\documentclass[11pt,a4paper]{article}
\pdfoutput=1
\usepackage{jcappub}
\usepackage{rotating}
\usepackage{array}
\usepackage{amsmath}
\usepackage[normalem]{ulem}
\usepackage{slashed}
\usepackage{bm}
\usepackage{booktabs}
\usepackage[pdftex,table]{xcolor}
\usepackage{units}

\newcommand{\xDM}{\mathbf{x}_\text{DM}}
\newcommand{\tprop}{\bm{\theta}_\text{prop}}
\newcommand{\ti}{\bm{\theta}_i}

\usepackage{multirow}

\newcommand{\diff}{\mathrm{d}}
\newcommand{\sv}{\langle \sigma v \rangle}

\newcommand{\pb}{{\bar{p}}}

\arxivnumber{TTK-21-28}

\title{Constraining dark matter annihilation
	with cosmic ray antiprotons using
	neural networks}

\author[1]{Felix Kahlhoefer,}
\author[2]{Michael Korsmeier,}
\author[1]{Michael Kr\"{a}mer,}
\author[1]{Silvia Manconi}
\author[1]{and Kathrin Nippel}

\affiliation[1]{Institute for Theoretical Particle Physics and Cosmology (TTK), RWTH Aachen University, D-52056 Aachen, Germany}
\affiliation[2]{The Oskar Klein Centre for Cosmoparticle Physics, Department of Physics, Stockholm University, Alba Nova, 10691 Stockholm, Sweden}

\emailAdd{kahlhoefer@physik.rwth-aachen.de}
\emailAdd{mkraemer@physik.rwth-aachen.de}
\emailAdd{michael.korsmeier@fysik.su.se}
\emailAdd{manconi@physik.rwth-aachen.de}
\emailAdd{nippel@physik.rwth-aachen.de}

\abstract{The interpretation of data from indirect detection experiments searching for dark matter annihilations requires computationally expensive simulations of cosmic-ray propagation. In this work we present a new method based on Recurrent Neural Networks that significantly accelerates simulations of secondary and dark matter Galactic cosmic ray antiprotons while achieving excellent accuracy. This approach allows for an efficient profiling or marginalisation over the nuisance parameters of a cosmic ray propagation model in order to perform parameter scans for a wide range of dark matter models. We identify importance sampling as particularly suitable for ensuring that the network is only evaluated in well-trained parameter regions. We present resulting constraints using the most recent AMS-02 antiproton data on several models of Weakly Interacting Massive Particles. The fully trained networks are released as \textsc{DarkRayNet} together with this work and achieve a speed-up of the runtime by at least two orders of magnitude compared to conventional approaches.}


\begin{document}
\maketitle
\flushbottom

\section{Introduction}
\label{sec:introduction}

The central prediction of the Weakly Interacting Massive Particles (WIMP)  paradigm is that Dark Matter (DM) particles should have a thermally averaged 
annihilation cross section of $\langle \sigma v \rangle \sim 10^{-26} \, \mathrm{cm^3 \, s^{-1}}$ during freeze-out. 
In many DM models, the present-day annihilation cross section in astrophysical systems is predicted to be of a similar magnitude, 
providing a clear target for indirect detection experiments searching for the products of DM annihilation processes.

While the most robust constraints on the DM annihilation cross section stem from observations of the {CMB~\cite{Aghanim:2018eyx} and of the} $\gamma$-ray sky, 
in particular from Fermi-LAT~\cite{Ackermann:2013yva,Ackermann:2015zua,Ackermann:2015lka}, highly complementary information
can be obtained by precisely measuring the flux 
of charged anti-particles arriving on Earth. Very recently, AMS-02 has released results 
from the first seven years of data taking~\cite{Aguilar:2021tos}, which include in particular the flux of antiprotons with unprecedented precision. 
Theoretical predictions for this flux however require detailed modelling of the production and propagation of 
charged cosmic rays (CRs) in the Galaxy, which are subject to significant uncertainties {and are currently constrained using CR data (see e.g.\ Ref.~\cite{Korsmeier:2021brc}), as well as their non-thermal emissions (see e.g.\ Ref.~\cite{Orlando:2017mvd}).} 

While various numerical codes, such as \textsc{Galprop}~\cite{Strong:1998fr} and \textsc{Dragon}~\cite{Evoli:2008dv}, exist to address this challenge and simulate the propagation of CRs, they require as input a large number of parameters that need to be varied to assess their impact on the predictions. 
As a result these simulations are typically computationally so expensive that they become prohibitive in the context 
of a global analysis of DM models, where also the fundamental model parameters need to be varied~\cite{Ambrogi:2018jqj}. Recent analyses of the AMS-02 antiproton data have therefore typically focused on simplified DM models with only a single annihilation channel, see e.g.\ Ref.~\cite{Cuoco:2016eej, Cui:2016ppb, Reinert:2017aga,Cholis:2019ejx}.

In the present work we explore the potential of artificial neural networks (ANNs) to solve this issue and substantially speed up 
the calculation of predictions for the primary antiproton flux for a very broad range of DM models.\footnote{For other recent works on the use of machine learning for cosmic ray propagation in the context of DM we refer to Refs.~\cite{Lin:2019ljc,Tsai:2020vcx}.} Specifically, we employ recurrent neural networks (RNNs), which are particularly well suited for the prediction of continuous spectra. 
The network is trained on a large sample of antiproton fluxes based on propagation parameters that are chosen to broadly agree with recent AMS-02 data, and a general parametrisation of the properties of the DM particle in terms of its mass and the branching fractions for various different final states. Using the same approach we have also developed and trained ANNs to accurately predict further CR species, like secondary antiprotons, protons or helium.

The predictions of the network can then be used to calculate the likelihood of the AMS-02 data for a given DM model and varying propagation parameters in order to calculate exclusion limits using a range of frequentist or Bayesian methods.  However, it is important to ensure that the resulting constraints are not biased by regions of parameter space for which the ANN has not been sufficiently trained. In the Bayesian approach this potential pitfall is avoided by evaluating the likelihood for a fixed sample of propagation parameter points drawn from the posterior probability distribution in the absence of a DM signal. The marginalisation over propagation uncertainties can then be performed via importance sampling, i.e.\ by appropriately reweighing and combining the points in the sample. This approach is particularly well suited for the analysis of antiproton data, since the propagation parameters are rather 
tightly constrained by the proton flux and the secondary antiproton flux, so that the presence of a DM signal 
does not dramatically shift the relevant regions of parameter space. 

We emphasise that, while the initial generation of a sample from the posterior is computationally expensive, it does not require specific assumptions on the properties of DM and therefore only needs to be carried out once in advance. Moreover, the same simulation step can be used to set up the training data for the ANN, ensuring that the network is trained specifically on the most interesting regions of parameter space. Once training is completed, the remaining steps are computationally cheap and can be performed for a large number of DM parameters. Indeed, the full marginalisation over propagation parameters can be performed in a similar amount of time as it would take to 
simulate a single parameter point in the conventional approach.

We apply our fully trained ANN to a number of cases of particular interest. For the case of DM annihilations exclusively into bottom quarks we show that the most recent AMS-02 data leads to results that are compatible with previous studies. 
In particular, we recover a notable excess for DM masses around 100 GeV in the case that no correlations in the AMS-02 data are considered. 
We also present new constraints on the well-studied model of scalar singlet DM and find that antiproton data places competitive constraints on this model. 
However, we emphasise  that the ANN is not limited to these cases and can be applied to a wide variety of DM models. 
Moreover, the general approach that we present can be extended to consider different propagation models (provided a suitable simulator exists), systematic uncertainties (such as correlations in the AMS-02 data) or cross-section uncertainties, enabling the community to fully explore the wealth of the available CR data.

The remainder of this work is structured as follows. In section~\ref{sec:cr} we briefly review the fundamental concepts of CR production and propagation and present the specific implementation that we adopt in the present work. We also carry out a first analysis of the most recent AMS-02 data and perform a parameter scan to identify the most interesting regions of parameter space. In section~\ref{sec:ANN} we introduce our machine learning approach to simulating CRs and discuss how we train and validate our ANNs. Finally, in section~\ref{sec:constraints} we apply the fully trained ANNs to constrain DM models. We present the relevant statistical methods and discuss the resulting exclusion limits.

\section{Cosmic-ray antiprotons in the Galaxy}\label{sec:cr}

For the following discussion it is useful to distinguish between primary and secondary CRs. Primary CRs are directly accelerated and emitted by standard astrophysical sources 
like supernova remnants or pulsars. But also more exotic scenarios such as the production of (anti)particles by DM annihilation or decay are considered as primary origin. 
Protons provide the dominant contribution to primary CRs (about 90\%) while helium (He) makes up about 10\%. Heavier nuclei only contribute at the percent level.
On the other hand, secondary CRs are produced during the propagation of primary CRs by fragmentation or decay.
More specifically, when the primary CRs interact with the gas in the Galactic disc, commonly called interstellar medium (ISM), secondary particles are produced.
Because of the different production mechanism, secondaries are suppressed with respect to primary CRs.
It is commonly believed that CR antiprotons do not have standard astrophysical sources%
\footnote{
	We note that the possibility of primary antiprotons that are directly produced and accelerated at supernova remnants
	\cite{Blasi:2009bd,Mertsch:2014poa,Mertsch:2020dcy, Kohri:2015mga} is also discussed in literature. 
}
such that their dominant contribution comes from secondary production. As a consequence, antiprotons are 
suppressed by 4--5 orders of magnitude with respect to protons, which makes them (together with other antimatter CRs, e.g.\ antideuterons \cite{Aramaki:2015pii,vonDoetinchem:2020vbj}) 
a promising channel for constraining DM signals. 

In this section we first discuss the production of antiprotons in the annihilation of dark matter particles in our Galaxy, followed by a discussion of backgrounds from secondary antiprotons. We then present the framework that we use to simulate CR propagation and the strategy to fit the resulting spectra to data. Finally, we perform a scan over the propagation parameters in order to create the training set for the machine learning approach introduced in section~\ref{sec:ANN}.

\subsection{Antiprotons from dark matter annihilation}\label{sec::antipdm}

CR antiprotons are a long standing target used to search for signals of WIMP DM in our Galaxy
\cite{Bergstrom:1999jc,Donato:2003xg,Bringmann:2006im,Donato:2008jk,Fornengo:2013xda,Evoli:2011id,Bringmann:2014lpa,Pettorino:2014sua,Cirelli:2014lwa,Cembranos:2014wza,Hooper:2014ysa,Boudaud:2014qra,Giesen:2015ufa,Evoli:2015vaa,Johannesson:2016rlh,Luque:2021ddh,DiMauro:2021qcf}.
More recently, there has been a discussion of an antiproton excess at about 20~GeV, which could be fitted with a 
primary DM source~\cite{Cuoco:2016eej, Cui:2016ppb, Reinert:2017aga, Cuoco:2019kuu, Cholis:2019ejx}. 
However, the excess might also be accounted for by a combination of systematic effects~\cite{Boudaud:2019efq, Heisig:2020nse, Heisig:2020jvs}.
If DM particles annihilate into standard model particle final states $f$ within the diffusion halo of our Galaxy as ${\rm DM}\!+\!{\rm DM} \to f\!+\!\bar{f}$, 
we expect a corresponding flux contribution to antiprotons in CRs, coming from the subsequent decay of for example $q \!+\!\bar{q}$ modes (see e.g.\ \cite{Cirelli:2010xx}). 
The source term of this primary antiproton component,  $q_{\bar{p}}^{(\mathrm{DM})}$, is a function of the Galactic coordinates $\bm{x}$ and the antiproton kinetic energy $E_\mathrm{kin}$. For a generic combination of standard model final states $f$ it reads: 
\begin{eqnarray}
  \label{eqn::pbar_DM_source_term}
  q_{\bar{p}}^{(\mathrm{DM})}(\bm{x}, E_\mathrm{kin}) =
      \frac{1}{2} \left( \frac{\rho(\bm{x})}{m_\mathrm{DM}}\right)^2  
      \sum_f \left\langle \sigma v \right\rangle_f \frac{\diff N^f_{\bar{p}}}{\diff E_\mathrm{kin}} \; .
\end{eqnarray}
The factor $1/2$ in eq.~\eqref{eqn::pbar_DM_source_term} corresponds to Majorana fermion DM.
Furthermore, $m_\mathrm{DM}$ is the DM mass, $\rho(\bm{x})$ the DM halo energy density profile, and $\sv_f$ is the thermally 
averaged annihilation cross section for the individual final states $f$. In the following, we fix $\sv$ independent of $f$ and account for this by assigning branching fractions into the relevant final states.
Finally, $\diff N^f_{\bar{p}}/\diff E_\mathrm{kin}$ denotes the energy spectrum of antiprotons for a single DM annihilation. 
This quantity depends on the DM mass and the standard model final state. Here we implement the  widely used tabulated results for the antiproton energy spectrum 
presented in Ref.~\cite{Cirelli:2010xx} which include electroweak corrections.%
\footnote{
	If DM annihilates into a pair of $W$ or $Z$ bosons it is possible to produce one of them off-shell. This possibility is not taken 
	into account in the original tables. We extend the tables of $W$ and $Z$ bosons to lower DM masses using the tables from Ref.~\cite{Cuoco:2017rxb}.
}

We assume that the DM density in our Galaxy follows an NFW profile~\cite{Navarro:1995iw}  
$\rho_{\mathrm{NFW}}(r) = \rho_h \, r_h/r\, \left( 1 + r/r_h \right)^{-2}$, 
with a scale radius of $r_h=20\;$kpc and a characteristic halo density, $\rho_h$, which is normalised such that the local 
DM density at the solar position of $8.5\;$kpc is fixed to $0.43\;$GeV/cm$^3$~\cite{Salucci:2010qr}, compatible also with more recent estimates \cite{deSalas:2020hbh}.
We note that the NFW profile is only one of many viable DM profiles currently investigated. Other profiles 
can have a significantly different behavior towards the Galactic center, see e.g.\ the discussion in Ref.~\cite{Benito:2019ngh}.
However, we stress that choosing a different DM density profile only has a small impact on the results presented in this paper  
since CR antiprotons from DM annihilation dominantly arrive from the local environment. 
Therefore they are mostly sensitive to the local DM density and the resulting flux depends only weakly on the shape of the DM density profile at 
the Galactic center. 
More specifically, the impact of changing the cuspy NFW profile to the cored Burkert profile~\cite{Burkert:1995yz} has been quantified 
in Ref.~\cite{Cuoco:2017iax}; it was found that a core radius of $5\;$kpc only weakens DM limits by about 20\%.

\subsection{Secondary antiprotons}\label{sec::sec}

The ISM consists of roughly  90\% hydrogen (H) and 10\% He. Thus secondary antiprotons are mostly produced by the 
interaction of $p$ and He CRs with the H and He components of the ISM. 
The source term for the secondary antiprotons $q_{\bar p}^{(\mathrm{sec})}$ is thus given by the convolution of the primary CR fluxes $\phi$ of isotope $i$, 
the ISM density $n_{\mathrm{ISM}}$ of component $j \in \lbrace \mathrm{H}, \mathrm{He} \rbrace$, and the energy-differential 
production cross section $\diff\sigma_{ij\rightarrow\bar p}/\diff  E_{\mathrm{kin},\bar{p}}$:
\begin{eqnarray}
	\label{eqn::pbar_sec_source_term}
	q_{\bar p}^{(\mathrm{sec})}({\bm x},E_{\mathrm{kin},\bar{p}}) &=&  
	                                \!\!\!\!\sum\limits_{j \in \lbrace \mathrm{H}, \mathrm{He} \rbrace} \!\!\!\! 4\pi \,n_{\mathrm{ISM},j}({\bm x}) 
	                                \sum\limits_{i} 
	                                \int 
	                                \diff E_{\mathrm{kin},i} \,
                                    \phi_i  ( E_{\mathrm{kin},i}) \, 
                                    \frac{\diff\sigma_{ij\rightarrow\bar p}}{\diff  E_{\mathrm{kin},\bar{p}} }(E_{\mathrm{kin},i} ,  E_{\mathrm{kin},\bar{p}} )\,.
\end{eqnarray}
By construction, secondaries are suppressed with respect to primary CRs. In the case of antiprotons, the experimentally 
observed suppression compared to protons is {5 orders of magnitude at 1~GV and increases to about 4 orders of magnitude above 10~GV}.
Since secondary CRs constitute the dominant contribution of the measured antiproton flux, 
considering standard astrophysical sources only already results in a good fit to the data \cite{Korsmeier:2016kha, Cuoco:2016eej, Boudaud:2019efq}, 
see also  discussion in section~\ref{sec:fitams}.

The cross section of secondary antiproton production is a very important ingredient of eq.~\eqref{eqn::pbar_sec_source_term}, which has been discussed 
by various groups recently~\cite{diMauro:2014zea,Winkler:2017xor,Korsmeier:2018gcy,Kachelriess:2019ifk}. 
In general there are two different strategies to determine this cross section. On the one hand, Monte Carlo generators, 
which are tuned to the relevant cross section data~\cite{Kachelriess:2019ifk}, can be used to infer the relevant cross section.
On the other hand, a parametrisation of the Lorentz invariant cross section can be fitted to all available cross section data. 
Then the required energy-differential cross section is obtained by an angular integration~\cite{diMauro:2014zea,Winkler:2017xor,Korsmeier:2018gcy}. 
We follow the second approach and use the analytic cross section parametrisation from Ref.~\cite{Winkler:2017xor} with the updated parameters from Ref.~\cite{Korsmeier:2018gcy}. 
An important advantage of the analytic cross section parametrisation is that it is explicitly tuned to cross-section data at low energies, and therefore more reliable 
below antiproton energies of $\sim 10$~GeV as discussed in Ref.~\cite{Donato:2017ywo}.

Finally, we consider that secondary antiprotons may scatter inelastically with the ISM and lose energy. This antiproton contribution, commonly referred to as tertiary \cite{Moskalenko:2001ya},
is suppressed with respect to the secondaries.

\subsection{Propagation in the Galaxy and solar modulation}\label{sec::prop}
The sources, acceleration and propagation of Galactic CRs are research topics by themselves \cite{Amato:2017dbs, Gabici:2019jvz}. 
Fast evolution and progresses has been driven in the last years by newly available and very precise data by 
AMS-02 \cite{Aguilar:2021tos}, PAMELA \cite{Adriani:2014xoa} and Voyager \cite{2013Sci...341..150S}. 
Some recent developments include the studies of systematic uncertainties from solar modulation, correlated experimental data points, 
secondary production/fragmentation cross sections as well as detailed studies of propagation phenomena below a rigidity of 10 GV to disentangle 
diffusion and reacceleration \cite{Genolini:2019ewc,Evoli:2019wwu,Evoli:2019iih,Boschini:2018baj,Boschini:2019gow,Weinrich:2020cmw,Weinrich:2020ftb,Luque:2021nxb,Luque:2021joz,Schroer:2021ojh, Korsmeier:2021brc}, {where the rigidity $R$ of a CR particle 
is defined as its momentum divided by the absolute value of its charge}.
Here we will not explore these exciting directions and instead focus on one standard setup of CR propagation, which was already studied in the context DM searches with antiprotons in Ref.~\cite{Cuoco:2019kuu}. The machine learning approach and the statistical methods introduced below can however be readily applied also to alternative assumptions and more refined descriptions. We briefly summarise below the main ingredients  of this specific approach and refer to Ref.~\cite{Cuoco:2019kuu} for a more detailed discussion.

\medskip

Charged CRs propagate within a diffusion halo assumed to be cylindrically symmetric, which extends a few kpc above and below the Galactic plane. 
In particular, it has a fixed radial extent of 20~kpc, while the 
half height of the diffusion halo is denoted by $z_\mathrm{h}$ and typically enters CRs fits as a free parameters (see section~\ref{sec:fitams}).
When CRs cross the boundary of the diffusion halo they escape from the Galaxy, while the propagation within the halo is described by a chain of coupled diffusion equations.

The diffusion equation for the {CR number density per volume and absolute momentum} 
$\psi_i (\bm{x}, p, t)$ of CR species $i$ at position $\bm{x}$ and momentum $p$  is given by~\cite{StrongMoskalenko_CR_rewview_2007}:
\begin{eqnarray}
  \label{eqn::propagationEquation}
  \frac{\partial \psi_i (\bm{x}, p, t)}{\partial t} = 
    q_i(\bm{x}, p) &+&  
    \bm{\nabla} \cdot \left(  D_{xx} \bm{\nabla} \psi_i - \bm{V} \psi_i \right)  \\ \nonumber
     &+&  \frac{\partial}{\partial p} p^2 D_{pp} \frac{\partial}{\partial p} \frac{1}{p^2} \psi_i - 
    \frac{\partial}{\partial p} \left( \frac{\diff p}{\diff t} \psi_i  
    - \frac{p}{3} (\bm{\nabla \cdot V}) \psi_i \right) -
    \frac{1}{\tau_{f,i}} \psi_i - \frac{1}{\tau_{r,i}} \psi_i \; .
\end{eqnarray}
We briefly describe each of the terms in eq.~\eqref{eqn::propagationEquation} below. 
To solve these equations numerically we employ \textsc{Galprop}~56.0.2870~\cite{Strong:1998fr,Strong:2015zva} and \textsc{Galtoollibs}~855\footnote{https://galprop.stanford.edu/download.php} 
	with a few custom modification as described in Ref.~\cite{Cuoco:2019kuu}. Alternatively, solutions might be obtained analytically, utilizing various simplifying assumption \cite{Putze:2010zn,Maurin:2018rmm}, or using other fully numerically codes like \textsc{Dragon}~\cite{Evoli:2008dv,Evoli:2017vim} or \textsc{Picard}~\cite{Kissmann:2014sia}.
\textsc{Galprop} assumes that CRs are in a steady state and solves the diffusion equations on a 3-dimensional grid.
Two dimensions describe the spatial distribution of CRs, the radial distance $r$ from the Galactic center and distance $z$ perpendicular to the plane, and 
one dimension contains the CR's kinetic energy. The grid points of the spatial dimensions are spaced linearly with step size of $\Delta r = 1$\;kpc 
and $\Delta z = 0.1$\;kpc, respectively, while the grid is spaced logarithmically in kinetic energy with a ratio between successive grid points of 1.4.

The source term $q_i$ in eq.~(\ref{eqn::propagationEquation}) depends on the CR species. For secondary antiprotons and antiprotons from DM annihilation it takes the form of 
eq.~\eqref{eqn::pbar_sec_source_term} and eq.~\eqref{eqn::pbar_DM_source_term}, respectively. 
For primary CRs the source term factorizes into a spatial and a rigidity-dependent term. 
The spatial term traces the distribution of supernova remnants.%
\footnote{
    We use the default prescription of \textsc{Galprop} where the parameters of the
    source term distribution are fixed to $\alpha = 0.5$, $\beta=2.2$, $r_s=8.5$~kpc, and $z_0=0.2$~kpc.
    This is slightly different from recent values in the literature \cite{Green:2015isa}.
    We note, however, that nuclei are only very weakly sensitive to the chosen distribution as discussed in Ref.~\cite{Korsmeier:2016kha}.
} 
On the other hand, the rigidity dependence is modeled as a smoothly broken power-law:
\begin{eqnarray}\label{eq::psp}
    \label{eqn::SourceTerm_2}
    q_R(R)     &=&   \left( \frac{R}{R_0} \right)^{-\gamma_1}
                     \left( \frac{R_0^{1/s}+R^{1/s} }
                                 {2\,R_0^{1/s}      } \right)^{-s (\gamma_2-\gamma_1)},
\end{eqnarray}
where $R_0$ is the break position and  $\gamma_{1,i}$ and $\gamma_{2,i}$ are the
spectral indices above and below the break for the CR species $i$, respectively.
The parameter $s$ regulates the amount of smoothing at the break.
{
     In the following analysis we will assume that all primary nuclei except for protons have a 
     universal injection spectrum such that we adopt $\gamma_{1,i}=\gamma_{1}$ and $\gamma_{2,i}=\gamma_{2}$. 
     For protons we allow different spectral behaviour and keep the subscript $i=p$.}
The broken power-law spectrum in eq.~(\ref{eq::psp}) is a widely used  phenomenological approximation which describes well the data in the considered rigidity range. 
All CR species are affected by several processes that contribute to CR propagation, which are diffusion, reacceleration, convection, and energy losses.
We assume that diffusion is spatially homogeneous and isotropic. In this case, the diffusion coefficient, $D_{xx}$, 
can be modeled as a broken power-law in rigidity
\begin{eqnarray}
    \label{eqn::diffusionConstant}
    D_{xx} &=&
        \begin{cases} 
       	    \beta D_{0} \left( \frac{R}{4 \, \mathrm{GV}} \right)^{\delta} &\text{if}\; R<R_1\\
       	    \beta D_{0} \left( \frac{R_1}{4 \, \mathrm{GV}} \right)^{\delta} \left( \frac{R}{R_1} \right)^{\delta_h} &\text{otherwise}\,,
        \end{cases}
\end{eqnarray}
where $D_0$ is an overall normalisation and $\delta$ and $\delta_h$ are the power-law indices below and above the break at position $R_1$.
At low energies the diffusion coefficient is proportional to the velocity $\beta=v/c$ of the CRs.
We allow for a diffusive reacceleration of CRs by scattering off Alfv\`en magnetic waves. The amount of reacceleration 
is then determined by the velocity $v_\mathrm{Alfven}$  of the waves  \cite{Ginzburg:1990sk,1994ApJ...431..705S}:
\begin{eqnarray}
    \label{eqn::DiffusivReaccelerationConstant}
    D_{pp} = \frac{4 \left(p \, v_\mathrm{Alfven} \right)^2 }{3(2-\delta)(2+\delta)(4-\delta)\, \delta \, D_{xx}} \; .
\end{eqnarray}

The terms proportional $\bm{V}(\bm{x})$ in eq.~(\ref{eqn::propagationEquation}) describe  convective winds which drive the CRs away from the Galactic plane. They are taken constant and orthogonal 
to the Galactic plane, such that $\bm{V}(\bm{x})= {\rm sign}(z)\, v_{0,{\rm c}}\,{\bm e}_z$.
The remaining terms describe different contributions of energy losses, for which we adopt the 
default \textsc{Galprop} implementation. In particular, continuous energy losses like ionisation or bremsstrahlung are 
included in the term $\diff p/\diff t$, while catastrophic energy losses by fragmentation or decay 
are modeled by the last two terms. The parameters $\tau_f$ and $\tau_r$ are the corresponding lifetimes. 

We emphasise again that the setup of CR propagation described above reflects one specific choice. The available measurements of 
CR nuclei are also described well by other setups. In particular, a model without diffusive re-acceleration,
but with an additional break in the diffusion coefficient between 5 and 10\;GeV 
is currently discussed in the literature \cite{Genolini:2019ewc,Weinrich:2020cmw,Korsmeier:2021brc}. 

\medskip

CRs measured at the top of the atmosphere  have to traverse a significant part of the heliosphere where they
are deflected and decelerated by solar winds. The strength of this effect varies in a 22-year cycle and is commonly 
known as solar modulation. It mostly affects low-energetic CRs; in practice the impact on the spectra above a few tens of 
GV is negligible. We use the common force-field approximation~\cite{Fisk:1976aw} to model the impact on the CR spectra:
\begin{eqnarray}
    \label{eqn::solarModulation}
    \phi_{\oplus,i}(E_{\oplus,i}) &=& \frac{E_{\oplus,i}^2 - m_i^2}{E_{\text{LIS}, i}^2 - m_i^2} \phi_{\text{LIS}, i}(E_{\text{LIS}, i}) \,,\\
    E_{\oplus,i} &=& E_{\text{LIS}, i} - e|Z_i|\varphi_{i}\,.
\end{eqnarray} 
{Here $\phi$ and $E$ label the energy-differential flux and the kinetic energy, respectively. 
The subscripts on the energy or flux denote the position which can either be local interstellar (LIS) or top of the atmosphere ($\oplus$). 
Furthermore, }$Z_i$ is the charge number, $e$ is the elementary charge, and $\varphi_{i}$ is the solar modulation potential. 
The potential is known to be time and charge-sign dependent.
We note that the force-field approximation is probably an oversimplified treatment of solar modulation.%
\footnote{
    In more sophisticated models solar modulation is described by propagation equations similar to eq.~\eqref{eqn::propagationEquation} but tuned to the 
    environment of the heliosphere. These are typically solved numerically~\cite{Kappl:2015hxv,Vittino:2017fuh,Boschini:2017gic,Fiandrini:2020puf,Ngobeni:2020quz}.
}
To minimise systematic impacts from solar modulation on our results we will exclude data below 5~GV from our analysis. Furthermore, 
we allow a different solar modulation potential for antiprotons  to account for a possible charge-sign dependence.  

\subsection{Fit to AMS-02 data}\label{sec:fitams}

In the following we summarise very briefly the considered data sets and the fit strategies, where the latter are directly adopted from Ref.~\cite{Cuoco:2019kuu}. 
The most precise measurement of CR antiprotons above 1~GV is currently provided by the AMS-02 experiment \cite{Aguilar:2021tos}. 
We consider the data sets of proton, helium, and {the antiproton-to-proton ratio} from AMS-02~\cite{Aguilar:2021tos} collected over 7 years from 2011 to 
2018 and complement with low-energy data for protons and helium from Voyager~\cite{Cummings:2016pdr}.
When fitting the CR data with the model outlined below, the CR likelihood is defined by 
\begin{eqnarray}
  	\label{eqn::likelihood_CR}
	-2\,\log{{\cal L}_{{\rm CR}} }(\bm\theta) =  \chi^2_{\rm CR}(\bm\theta) =  \sum\limits_{e,s,i} 	
							\left(\frac{\phi^{(e)}_{{ s},i}- \phi^{(\text{m})}_{s,i,e}(\bm \theta)}
							           {\sigma^{\left(e\right)}_{s,i}}\right)^2 \,,
\end{eqnarray}
where $\phi^{(e)}_{{s},i}$ denotes the flux of the CR species $s$ that was measured by the experiment $e$
at the rigidity $R_i$ or energy $E_{\mathrm{kin},i}$, 
while $\phi^{(\text{m})}_{s,i,e}(\theta)$ is the flux computed with \textsc{Galprop}  for the corresponding species and energy. 
Finally, $\sigma^{\left(e\right)}_{s,i}$ is the experimental uncertainty of the flux measurement.
The AMS-02 experiment provides separate statistical and systematic uncertainties. Here we assume that 
the systematic uncertainties are uncorrelated and add the two contribution in quadrature. This 
is certainly a simplified treatment.
In particular, it was shown that the significance of a potential excess can depend critically on the assumptions made for  the correlations of this
uncertainty~\cite{Boudaud:2019efq,Heisig:2020nse}. 
However, we expect that the impact on DM limits is less severe,
{
    because of two opposing effects: The covariance matrices as modeled in Refs.~\cite{Boudaud:2019efq,Heisig:2020nse}
    contain contributions with both large and small correlation lengths. A large correlation length corresponds 
    to a change of the overall normalisation which looks very different from a peaked signature 
    such as expected from DM annihilation. This potentially leads to 
    a stronger DM limit. On the other hand, a small correlation length allows for signatures similar to those from DM and, therefore, potentially weakens the limit.
    Overall, we expect these two effects to partly cancel each other. 
}
We leave the study of different systematics and correlations within the new methods introduced in this paper to future investigation.

Within the phenomenological description of CR injection and propagation outlined in section~\ref{sec::prop}, the parameters of 
eq.~\eqref{eqn::propagationEquation} are largely unconstrained a priori and are directly inferred from CR data. 
We allow for a total of 15 parameters to describe CR injection and propagation. To sample this large parameter space efficiently we use the Nested Sampling algorithm implemented in the \textsc{MultiNest} code~\cite{Feroz:2008xx}. The computing efficiency is increased even further by exploiting a hybrid strategy where only a subset of parameters 	is sampled by \textsc{MultiNest} (``slow parameters'') and the remaining parameters are profiled on-the-fly (``fast parameters''). 	The slow parameters are the ones that are needed as input for \textsc{Galprop} and thus changing them is time consuming. More specifically, these are the following eleven parameters: the slopes of the primary injection spectra $\gamma_{1,p}$, $\gamma_{1}$, $\gamma_{2,p}$, 
and $\gamma_{2}$, the break position $R_0$ and its smoothing $s$, the normalisation $D_0$ and slope
$\delta$ of the diffusion coefficient, the half-height of the diffusion halo $z_\mathrm{h}$, and the velocities of
convection $v_{0c}$ and Alfv{\`e}n magnetic waves  $v_\text{Alfv{\`e}n}$.
The scan ranges for all of these parameters are summarised in table~\ref{tab:param_ranges_v2}.
{In the following we will give results in the frequentist and the Bayesian interpretation. 
For the Bayesian interpretation we assume flat priors in the scan ranges.}

The four remaining parameters describe the normalisation of the proton ($A_p$) and helium ($A_\mathrm{He}$) fluxes and the solar modulation potentials ($\varphi_{\text{AMS-02},p,{\rm He}}$ for $p$ and He and $\varphi_{\text{AMS-02},\pb}$ for  $\pb$). These are the fast parameters, which are treated in a simplified way in our analysis and therefore can be varied much more easily.
Instead of explicitly including them in the \textsc{MultiNest} parameter scans, we profile over them on-the-fly at each likelihood evaluation of \textsc{MultiNest}, i.e.\ we maximise the likelihood over the fast parameters using \textsc{Minuit}~\cite{James:1975dr}. A very weak Gaussian prior is applied to $\varphi_{\text{AMS-02},\pb}$ by adding to the main likelihood the term
$-2\,\log({{\cal L}_{{\rm SM}} }) = (\varphi_{\text{AMS-02},p,{\rm He}}-\varphi_{\text{AMS-02},\pb})^2 / \sigma_\varphi^2$
where $\sigma_\varphi = 100$ MV,%
\footnote{
	 {The prior expresses that the solar modulation potential of antiprotons 
     and the one of protons and helium are related, even if they are not forced to be the same.
     In Ref.~\cite{Vittino:2019yme} the average difference between the potential of electrons and positrons 
     was found to be around 100 MV.}
}
while no priors are applied on $\varphi_{\text{AMS-02},p,{\rm He}}$. 

We truncate the rigidity range of the fit to the range between 5 to 300\;GV. 
As mentioned above, data below 5\;GV is 
excluded to avoid a strong bias from our modeling of solar modulation.%
\footnote{
    It was shown in Ref.~\cite{Cuoco:2019kuu} that the cut at $R=5$\;GV does not artificially enhance the significance of a potential DM signal.  
}
At high energies, the spectra of CR nuclei show a break at $R\sim300$\;GV, which is more pronounced 
in secondaries with respect to primaries~\cite{Aguilar:2018njt,Aguilar:2017hno}.
While in general it would be possible to introduce spectral breaks in the injection spectrum or in the diffusion coefficient, 
only the latter naturally explains the different behavior of the primaries and secondaries~\cite{Genolini:2017dfb}. We therefore 
fix the parameters of eq.~\eqref{eqn::diffusionConstant} to $R_1=300$ GV and $\delta_h=\delta - 0.12$. The proton and helium data 
of AMS-02 is described well by this choice. Truncating our fit at $R\sim300$\;GV avoids unnecessary bias.%
\footnote{
    As an alternative, $R_1$ and $\delta_h-\delta$ could be treated at free parameters in the fit which would, however, increase the 
    complexity of the already high-dimensional parameter fit. 
}

\begin{table}[t]
	\centering
	\renewcommand{\arraystretch}{1.3}
	\caption{
        Results of the CR fits to AMS-02 and Voyager data of protons, helium, and antiprotons. The parameter ranges for the \textsc{MultiNest} scan are stated in column 2. 
        In the remaining columns we state the best-fit parameter values and their uncertainty at the 68\% C.L. for a fit with and without a DM signal. 
        Results are given both in the frequentist and the Bayesian interpretation. 
        \label{tab:param_ranges_v2}
	 }
	\begin{tabular}{lccccc}
		\hline
		\hline
		                                                                    &               &  \multicolumn{2}{c}{Frequentist}                                                        & \multicolumn{2}{c}{Bayesian}                                                            \\
		Parameter                                                           &  Scan ranges  & w/o DM                                     & w/ DM                                      & w/o DM                                     & w/ DM                                      \\
		\hline
		$\gamma_{1,p}$                                                      & $[1.2, 2]   $ & $                  {1.80}^{+0.04}_{-0.03}$ & $                  {1.79}^{+0.07}_{-0.06}$ & $                  {1.77}^{+0.07}_{-0.04}$ & $                  {1.68}^{+0.14}_{-0.07}$ \\
		$\gamma_{1}$                                                        & $[1.2, 2]   $ & $                  {1.79}^{+0.04}_{-0.04}$ & $                  {1.74}^{+0.08}_{-0.06}$ & $                  {1.75}^{+0.07}_{-0.04}$ & $                  {1.63}^{+0.15}_{-0.07}$ \\
		$\gamma_{2,p}$                                                      & $[2.3, 2.6] $ & $               {2.405}^{+0.013}_{-0.007}$ & $                  {2.48}^{+0.02}_{-0.03}$ & $                  {2.41}^{+0.01}_{-0.01}$ & $                  {2.48}^{+0.02}_{-0.03}$ \\
		$\gamma_{2}$                                                        & $[2.3, 2.6] $ & $               {2.357}^{+0.014}_{-0.005}$ & $                  {2.42}^{+0.02}_{-0.03}$ & $               {2.366}^{+0.009}_{-0.012}$ & $                  {2.42}^{+0.02}_{-0.02}$ \\
		$R_{0}\,\mathrm{[ 10^{3}\;MV]}$                                     & $[1, 20]    $ & $                  {7.92}^{+0.82}_{-0.80}$ & $                  {7.32}^{+1.16}_{-0.83}$ & $                  {7.06}^{+0.93}_{-1.04}$ & $                  {6.42}^{+0.97}_{-1.13}$ \\
		$s_{}$                                                              & $[0.1, 0.9] $ & $                  {0.37}^{+0.03}_{-0.03}$ & $                  {0.40}^{+0.03}_{-0.04}$ & $                  {0.38}^{+0.04}_{-0.04}$ & $                  {0.44}^{+0.04}_{-0.06}$ \\
		$D_{0}\,\mathrm{[ 10^{28}\;cm^2/s]}$                                & $[0.5, 10]  $ & $                  {2.05}^{+1.48}_{-0.39}$ & $                  {2.92}^{+2.09}_{-0.96}$ & $                  {3.58}^{+1.30}_{-0.73}$ & $                  {5.37}^{+1.52}_{-1.78}$ \\
		$\delta$                                                      & $[0.2, 0.6] $ & $               {0.419}^{+0.009}_{-0.012}$ & $                  {0.35}^{+0.03}_{-0.02}$ & $                  {0.42}^{+0.01}_{-0.01}$ & $                  {0.33}^{+0.03}_{-0.03}$ \\
		$v_\mathrm{Alfven}\,\mathrm{[km/s]}$                                & $[0, 30]    $ & $                  {8.84}^{+1.45}_{-2.58}$ & $                 {10.25}^{+2.12}_{-2.06}$ & $                  {6.02}^{+3.57}_{-2.51}$ & $                  {7.70}^{+4.15}_{-3.10}$ \\
		$v_{0,\mathrm{c}}\,\mathrm{[km/s]}$                                 & $[0, 60]    $ & $                  {0.09}^{+1.08}_{-0.08}$ & $                  {0.90}^{+6.77}_{-0.78}$ & $                  {2.48}^{+0.32}_{-2.48}$ & $                {13.36}^{+2.44}_{-13.36}$ \\
		$z_\mathrm{h}\,\mathrm{[kpc]}$                                      & $[2, 7]     $ & $                  {2.60}^{+2.25}_{-0.48}$ & $                  {2.79}^{+2.87}_{-0.75}$ & $                  {4.70}^{+2.30}_{-0.86}$ & $                  {4.84}^{+2.13}_{-0.75}$ \\
		$\log_{10}(m_\mathrm{DM}/\mathrm{MeV})$                          & $[4, 7]     $ & -                                          & $                  {5.07}^{+0.03}_{-0.05}$ & -                                          & $                  {5.08}^{+0.04}_{-0.05}$ \\
		$\log_{10}( {\langle} \sigma v {\rangle} \mathrm{s/cm^{3}})$     & $[-27,-22]  $ & -                                          & $                {-25.42}^{+0.22}_{-0.48}$ & -                                          & $                {-25.76}^{+0.13}_{-0.26}$ \\
		$\varphi_{\mathrm{AMS-02,pHe}}\,\mathrm{[GV]}$                      &               & $                  {0.26}^{+0.04}_{-0.03}$ & $                  {0.25}^{+0.05}_{-0.03}$ & $                  {0.30}^{+0.04}_{-0.05}$ & $                  {0.28}^{+0.04}_{-0.06}$ \\
		$(\varphi_{\bar p} - \varphi_{p})_\mathrm{AMS-02} \,\mathrm{[GV]}$  &               & $               {0.200}^{+0.000}_{-0.036}$ & $                  {0.13}^{+0.07}_{-0.12}$ & $               {0.177}^{+0.023}_{-0.001}$ & $                  {0.09}^{+0.11}_{-0.03}$ \\
		$A_{\mathrm{p,AMS-02}}$                                             &               & $               {1.173}^{+0.004}_{-0.003}$ & $               {1.173}^{+0.003}_{-0.004}$ & $               {1.178}^{+0.004}_{-0.004}$ & $               {1.177}^{+0.004}_{-0.004}$ \\
		$A_{\mathrm{He,AMS-02}}$                                            &               & $               {1.257}^{+0.006}_{-0.014}$ & $                  {1.20}^{+0.02}_{-0.01}$ & $               {1.253}^{+0.010}_{-0.010}$ & $                  {1.20}^{+0.02}_{-0.02}$ \\
		$\chi^2_{\mathrm{p,AMS-02}}$                                        &               & $                                     7.2$ & $                                     6.2$ \\
		$\chi^2_{\mathrm{He,AMS-02}}$                                       &               & $                                     3.2$ & $                                     2.1$ \\
		$\chi^2_{\mathrm{pbar/p,AMS-02}}$                                   &               & $                                    35.0$ & $                                    21.5$ \\
		$\chi^2_{\mathrm{p,Voyager}}$                                       &               & $                                     7.9$ & $                                     4.1$ \\
		$\chi^2_{\mathrm{He,Voyager}}$                                      &               & $                                     3.9$ & $                                     3.2$ \\
		$\chi^2$                                                            &               & $                                    57.2$ & $                                    37.1$ \\
		\hline \hline
	\end{tabular}
	\renewcommand{\arraystretch}{1.0}
\end{table}

To gain a first understanding of the allowed regions of parameter space, we perform a CR fit as detailed above. The scan is conducted using 1000 live points, a stopping criterion of \textsc{tol=0.1} and an enlargement factor \textsc{efr=0.7}. 
The final efficiency of the scan is found to be around 9\%, 
with about 350\,000 
likelihood evaluations in total.
     The fit is heavily parallelised using 96 cores at the same time. The \textsc{Galprop} code is parallelised using \textsc{openMP} while the nested sampling algorithm of
     \textsc{MultiNest} can be expanded to multiple \textsc{MPI} tasks. We follow a hybrid strategy with 24 \textsc{MPI} tasks using 4 cores for each task. 
     We have verified that the parallelisation efficiency lies above 70\%. 
     In total the fit requires about 5.5 days to converge and consumes 12500 cpu hours, which means that a single likelihood evaluation requires on average about 130 cpu seconds. 
To perform this fit, \textsc{MultiNest} starts by broadly sampling the entire parameter space and then continuously shrinks to the allowed parameters. The result is an ensemble of parameter points which is denser in the most interesting parameter region. We will make use of this property in the following section, where our goal is to train the ANNs in such a way that they perform particularly well in the parameter range preferred by data. Thus, we save all the sample points during the fit and then use them as a starting point for the training in section~\ref{sec:ANN}.

In table~\ref{tab:param_ranges_v2} we summarise the best-fit (most probable) values of the various parameters based on a frequentist (Bayesian) interpretation as well as their 68\% confidence intervals (credible intervals).
The best-fit point corresponds to $\chi^2 = 57.2$ for the AMS-02 data. 
These results broadly agree with the ones from Ref.~\cite{Cuoco:2019kuu} even though we use the more recent 7-year AMS-02 data for $p$, He, and $\bar p$.
As expected, for the parameters that are well-constrained by data, there is good agreement between the frequentist and the Bayesian approach. For less constrained parameters (such as for example $z_\mathrm{h}$) there can be sizeable differences between the best-fit point (obtained by maximizing the profile likelihood) and the most probable point (obtained by maximizing the marginalised likelihood). We will return to this issue in section~\ref{sec:constraints}.

Previous analysis have discussed a potential DM signal that could be accommodated at antiproton energies between 10 and 20 GeV where the antiproton flux shows a small anomaly at the level of a few percent. This potential signal corresponds, for example, to DM particles with a mass of about 80~GeV that self-annihilate into $b\bar{b}$ final states at a thermal cross section. However, the significance of this potential signal has been discussed controversially in the literature. The most recent works suggest that the anomaly is well explained by the combination of several systematic uncertainties, namely uncertainties in the secondary antiproton production cross section, correlated systematics in the AMS-02 data, and some additional 
freedom in the CR propagation model \cite{Boudaud:2019efq, Heisig:2020nse, Heisig:2020jvs}, which we do not include here. 

The focus of this work lies instead on developing new methods for exploiting ANNs and importance sampling to derive DM limits. In contrast to a DM signal, we expect the limits to be only weakly dependent on those systematics and leave their investigation to future studies. Nevertheless, for comparison we also perform one fit where antiprotons from DM annihilation are included. We choose DM annihilation into a pair of $b\bar b$ quarks as our benchmark. In this case,
two further parameters are considered, the mass of the DM particle $m_{\rm DM}$ and the thermally averaged annihilation cross section $\sv$ (see eq.~\eqref{eqn::pbar_sec_source_term}). We explore values of $m_{\rm DM}$ from 10\;GeV to 10\;TeV and values of $\sv$ between $10^{-22}$ and $10^{-27}\;\mathrm{cm^3/s}$, with our results being independent of the precise choice of these ranges.
Theses two additional parameters are sampled with \textsc{MultiNest}. 

The results of the additional fit are also shown in table~\ref{tab:param_ranges_v2}.
Including a DM signal formally improves the $\chi^2$ by 20.1 which, however, 
given the discussion above should not be interpreted as significant. Nonetheless, we can take this value as a point of 
comparison for the performance of the ANN in section~\ref{sec:constraints}. 
{We furthermore observe that, while the additional DM signal affects most CR propagation parameters only marginally, there is a sizeable shift of the preferred parameter regions for $\gamma_{2}$, $\gamma_{2,p}$ and $\delta$. While this shift is likely overestimated in our analysis for the reasons mentioned earlier, it highlights the challenges for the training of the ANN (see section~\ref{sec:ANN}) and for the statistical inference via importance sampling (see section~\ref{sec:constraints}).}

In the previous paragraph, we focused on a specific case of DM annihilation into a pair of bottom quarks which serves as an example and a point of comparison. In general,
much more complex scenarios with a range of different final states and combinations at different branching fractions are possible. The naive approach to obtain
results would be to perform an entirely new parameter scan for each case of interest which obviously requires a substantial amount of computational resources. 
Instead, in the following we will discuss methods to speed up the calculation of CR spectra in a model-independent fashion to quickly obtain constraints 
for any given DM model.

\section{Deep neural network setup and training}
\label{sec:ANN}

Our aim is to predict the output of \textsc{Galprop} for a wide range of input parameters representing both uncertainties in the propagation model and the unknown properties of DM. This output can then be used to calculate experimental likelihoods as described in section~\ref{sec:fitams} without computationally expensive simulations. To achieve this goal, we build and train suitable ANNs and validate their performance. Considering the two different contributions to the antiproton flux (i.e.\ primary and secondary CRs), we construct two separate ANNs to provide fast predictions of each component based on the relevant physical parameters. 
We will refer to the networks for the DM component and the secondary component as DMNet and sNet, respectively.
As the underlying method in the development of the neural networks is the same, both ANNs will be presented in parallel in this section. 

\subsection{Training Set}
\label{sec:training_set}

The information that a neural network should be able to learn, in general, has to be represented in the data that is used to train the network. This allows for the interpolation of data within the parameter space that would, in a conventional approach, require new simulations. To remain impartial on the specific parameters of the DM model, we consider a wide range in the mass of the DM particle from 5 GeV to 5 TeV and randomly sample from a logarithmic distribution in this range. A similar approach is taken for the branching fractions, where we consider all SM final states that give a non-negligible contribution to a CR antiproton flux~\cite{Cirelli:2010xx}: $q\bar{q}$, $c\bar{c}$, $b\bar{b}$, $t\bar{t}$, $W^+W^-$, $ZZ$, $hh$ and $gg$. We logarithmically sample each branching fractions in the range $[10^{-5}, 1]$ and then normalise the result in such a way that the sum of all  branching fractions equals one. The DM annihilation cross section is fixed to $\langle \sigma v \rangle = 3 \times 10^{-26} \, \text{cm}^3 \, \text{s}^{-1}$ in the complete training set, as variations in this parameter can be included at a later stage by an appropriate rescaling of the flux. 
These DM parameters, which we will collectively denote by $\xDM$, are only relevant to the DM component of the antiproton flux and the corresponding neural network, while the secondary flux is independent of $\xDM$ and hence these parameters will not be used as inputs to the sNet. 

\begin{figure}[t]
	\centering
	\includegraphics[width = \textwidth,clip,trim=5 0 5 0]{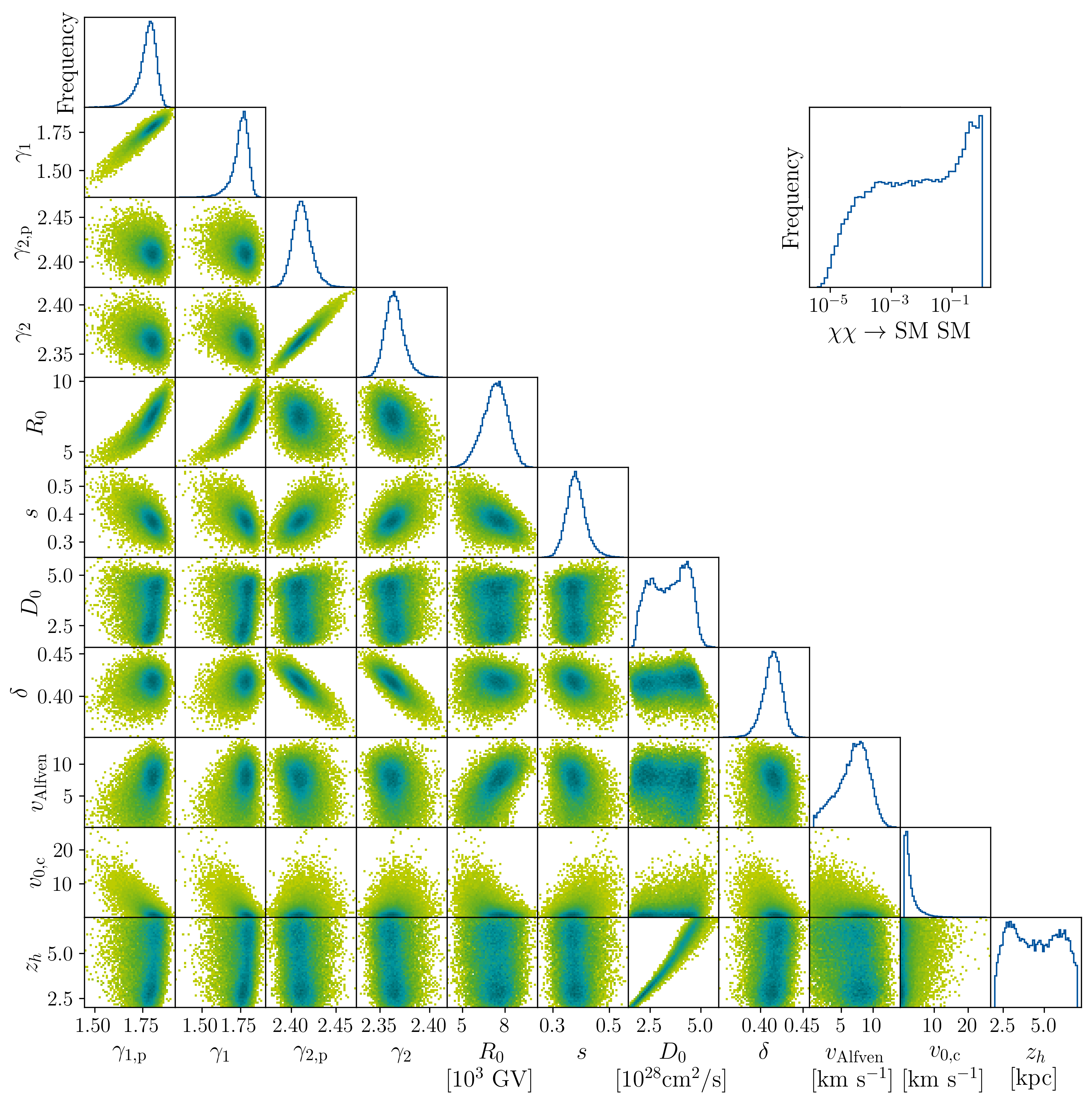}
	\caption{\textit{Triangle plot:} One and two dimensional histograms {showing the frequency} of propagation parameters used in the training set, constructed in such a way that the highest density is achieved in the regions most favoured by the combination of AMS-02 proton, antiproton and helium data without DM signal. \textit{Top right:} One dimensional histogram of the training set for each of the branching fractions {$\chi \chi \rightarrow$ SM SM}.}
	\label{img:training_data_distibution}
\end{figure}

For the propagation parameters we face the significant challenge that the parameter space introduced in section~\ref{sec:cr} is 11-dimensional and that only a very small volume of this parameter space gives an acceptable fit to AMS-02 data. If we were to simply perform a grid scan or draw random samples from this parameter space, we would include large regions of parameter space for which accurate predictions are unnecessary, as they will anyways be strongly excluded. Conversely, in the preferred regions of parameter space, we want to achieve an accuracy that is significantly better than the typical relative errors of about 5\% in AMS-02 data, which requires large amounts of training data.

To obtain sufficiently accurate network predictions in the most interesting regions of parameter space without spending unnecessary time on the simulation and training of less interesting regions, we want to make use of the AMS-02 data already for the creation of the training set. Indeed, we can directly use the \textsc{MultiNest} scan described in section~\ref{sec:fitams} to obtain a sample of propagation parameters (denoted by $\tprop$ in the following) that is focused on the regions of parameter space with the highest likelihood (see also Ref.~\cite{Coccaro:2019lgs}). {Since in the following we will be most interested in the calculation of exclusion limits, we will base our training on the \textsc{MultiNest} scan without DM signal. For a detailed investigation of the excess the same procedure outlined below could be applied to the sample of propagation parameters from the \textsc{MultiNest} scan with DM signal.}

For the creation of the training set we exclude any parameter point in the \textsc{MultiNest} sample that gives a poor fit to AMS-02 data, specifically with $\Delta \chi^2 \ge 30$ compared to the best-fit point. This results in a total of 117335 remaining parameter points which we show in figure~\ref{img:training_data_distibution}. We emphasise that for each parameter the training data extends well beyond the 68\% confidence/credible intervals {without DM annihilations quoted in table~\ref{tab:param_ranges_v2}}. To ensure a sufficiently good coverage also of the DM parameter space, we sample 8 combinations of DM parameters for each propagation parameter point, leading to a very large simulation set of $\mathcal{O} (10^6)$ parameter points. 

\subsection{Neural Network Architectures}
\label{sec:architectures}

Although the two networks that we use to predict the two components of the antiproton flux can be set up and trained in a similar way, we face distinct challenges in each component. For the DMNet the key challenge is the very large number of input parameters, namely the DM mass plus 8 branching fractions in $\xDM$ and a total of 11 propagation parameters in $\tprop$, each with a different physical effect on the output, i.e.\ the antiproton flux. As we want to have accurate predictions for variations in each of the parameters, we treat the DM mass, the branching fractions, and the propagation parameters as three distinct sets of inputs, which are first processed by three independent dense networks before combining the outputs (see below).

For the sNet the key challenge is to achieve sufficient accuracy in the prediction of the secondary antiprotons flux, which is tightly constrained by AMS-02 data. Given these constraints, the secondary antiproton flux only exhibits relatively small variations across the training set, which nevertheless need to be accurately captured using the sNet. To achieve the desired accuracy, we provide an increased number of trainable parameters that define the network.
As we will show within the following sections, the training duration consequently increases with respect to the DMNet but a very good accuracy is achieved.

Rather than directly feeding the physical parameters as inputs to the network, we map the logarithm of $\xDM$ to values in the range $\left[0, 1\right]$ and the remaining parameters $\tprop$ to a distribution with a mean of 0 and a standard deviation of 1. Each of the networks is then trained in a supervised approach. The simulated fluxes serve as training labels or `true' fluxes to which the network output can be compared.

Given the large variations in the CR fluxes that are desired as the output of the ANNs, here we choose a natural scaling of the original (simulated) flux $\Phi ( E )$ for the sNet outputs, 
\begin{equation}
	\tilde{\Phi}_\text{s} (E) = \log_{10} \left( \Phi(E) \, E^{2.7} \right) \, .
	\label{eq:trafo}
\end{equation}
The $\log_{10}$ further decreases the variations in the flux values, which would otherwise cover several orders of magnitude.
The energies and their respective fluxes are binned values, identical to the output from the simulations, which extend over the energy range of the AMS-02 antiproton measurement. Consequently, we have sequences of distinct values in the scaled flux as training labels. The transformation in eq.~(\ref{eq:trafo}) is easily invertible and thus allows for direct comparison of the network output to the simulated spectra.

\begin{figure}[t]
	\centering
	\includegraphics[width = 0.7\textwidth]{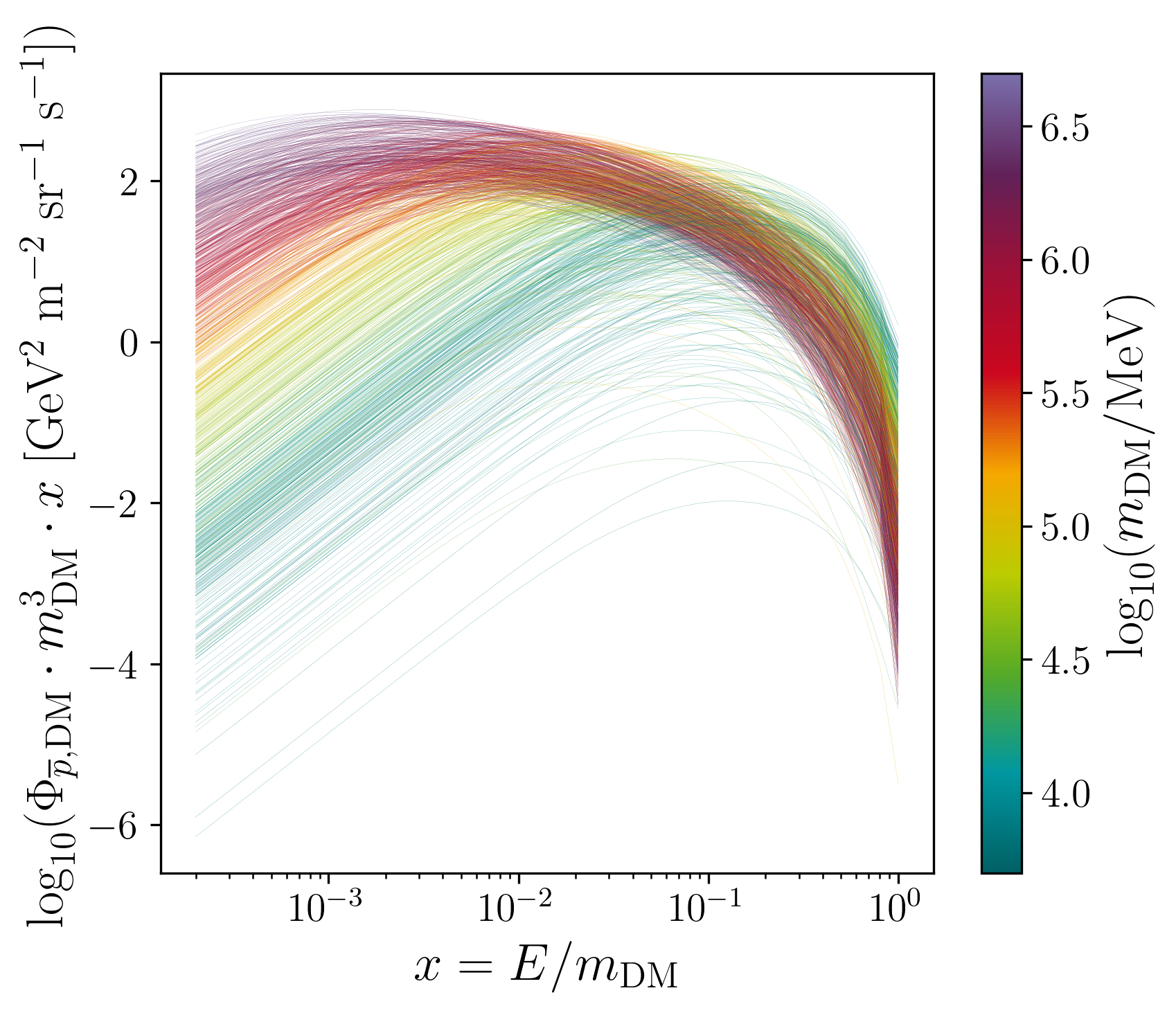}
	\caption{Transformed DM antiproton fluxes following eq.~\eqref{eq:trafo_DM} for our training set, which varies propagation parameters, branching fractions and DM masses as discussed in sec.~\ref{sec:training_set}. The modest amount of variation across different parameter points results in a more easily processable version of the input \textsc{GALPROP} simulated flux for the DMNet.} 
	\label{img:training_outputs}
\end{figure}

As the DM component of the flux predominantly scales with the DM mass, we choose a different scaling for that flux component, 
\begin{equation}
	\tilde{\Phi}_\text{DM} (x) = \log_{10} \left( m_\text{DM}^3 \, x \, \Phi(E) \right) \, ,
	\label{eq:trafo_DM}
\end{equation}
where $x = E/m_\text{DM}$ is a dimensionless quantity. We use a grid in $x$ with 40 points logarithmically spaced in the interval $[10^{-3.7},1]$, on which we evaluate the training labels and DMNet output. The advantage of this scaling compared to eq.~\eqref{eq:trafo} is that it substantially reduces the impact of changing the DM mass and therefore leads to much less variation across the training set.\footnote{To first approximation the DM component of the antiproton flux follows the source term $
\Phi_{\overline{p}, \text{DM}} \left(E \right) \propto q_\text{DM} \propto m_\text{DM}^{-2} \, \mathrm{d} N/\mathrm{d} E \propto m_\text{DM}^{-3} x^{-1} \, \mathrm{d} N/ \mathrm{d} \log_{10} x$, where $\mathrm{d} N / \mathrm{d} \log_{10} x$ depends only very mildly on $m_\text{DM}$.
}
This is illustrated in figure~\ref{img:training_outputs}, which shows the resulting DM antiproton fluxes $\tilde{\Phi}_\text{DM}$ as a function of $x$ for a representative set of final state combinations and DM masses in the training set. We find that for each combination of input parameters we obtain a slowly-varying function of $x$ that reaches a maximum and then drops towards $x \to 1$. The general trend is similar across the entire range of DM masses that we consider, but some information on the DM mass is retained. We find that this approach significantly improves the training of the DMNet compared to the scaling in eq.~\eqref{eq:trafo}.

\begin{figure}[t]
	\centering
	\includegraphics[width = 1\textwidth]{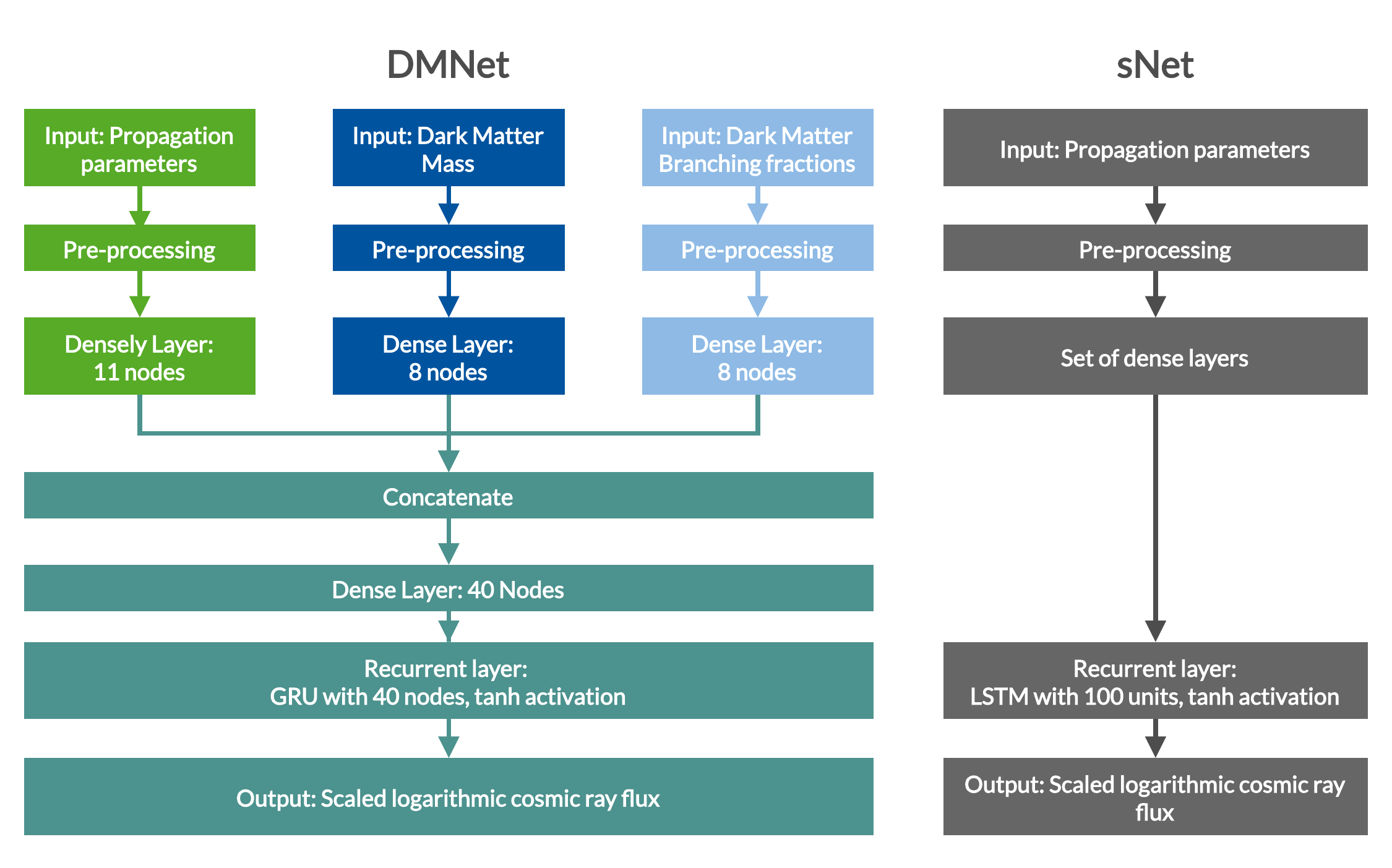}
		\begin{tabular}{ll}
            \hline \hline
			\multicolumn{2}{l}{\textbf{Hyperparameters}} \\
			\hline
			Activation & ReLU \\ 
			Dropout fraction & 0.1 \% \\ 
			Optimizer & Adam, learning rate scheduling $l \in [10^{-2}, 10^{-5}]$, patience 10 epochs \\ 
			Loss & Mean squared error (MSE) \\
			Batch size & 500 \\ 
			Validation split & 20~\% \\
			Early stopping & Monitor val. loss, patience = 40 \\
			\hline
			\hline
		\end{tabular}
	\caption{Schematic of the network structure. \textit{Top left:} Architecture set up for handling the complete set of inputs. This network type can be used to be trained on the DM component of the $\overline{p}$ flux (DMNet). \textit{Top right:} Simplified architecture for networks that require only the CR propagation parameters as input. This network architecture is designed for learning the $\overline{p}_\text{secondary}$ fluxes (sNet) and can be employed to train on proton and helium spectra as well (see appendix~\ref{app:p_and_He}). \textit{Bottom:} The hyperparameters used during the training process for each of these networks.}
	\label{img:architecture}
\end{figure}

Subsequent to the pre-processing of the input, the ANNs contain densely connected ('dense') layers, that process the information from the inputs. To address the individual challenges for the networks we set up the architecture as depicted in figure~\ref{img:architecture}. We provide dense layers for each of the different inputs in the DMNet which are concatenated in the next step and followed by large dense layers. In the sNet the pre-processed input is fed through a more intricate set of dense layer, specifically (56, 28, 14, 28, 56) nodes in the set of layers. We use ReLU activations and add a small dropout probability of 0.1\% between the layers. The precise values of these hyperparameters do not significantly affect the training performance.

The main feature of each of the networks is a recurrent layer. The choice to work with a recurrent setup instead of other network architecture types has lead to significant improvements in the architecture development process. Even though the typical application for RNNs is time-series data, we find our spectra as functions of energy to be handled just as well by this network type. In particular, it can be reasoned that the information on the flux that is contained in a specific energy bin is highly correlated with the prior and subsequent energy bins and a network architecture that is able to propagate the information of neighbouring units is very beneficial for the task at hand.
We chose a GRU layer as proposed in Ref.~\cite{DNN:gru} in the DMNet and a LSTM layer following \cite{DNN:lstm} in the sNet. Each of these layer types is useful for long data sequences and far-reaching information propagation without leading to vanishing or exploding gradients during training. While both methods can in principle be used for either networks, the final implementations that achieved the best results was based on different layer types.
As network output a final dense layer is set up. We build the networks using the deep learning API \textsc{Keras} \cite{DNN:keras} which uses \textsc{Tensorflow} \cite{DNN:tensorflow} as backend. 

\renewcommand{\arraystretch}{1.5}

\subsection{Training process}
\label{sec:train_process}

We use approximately $75$\% of the previously described simulation set for the network training.\footnote{Note that the sNet has a smaller training set compared to the DMNet, as here we have fewer unique spectra following from our parameter sampling for simulating the training set.} The remainder is used as a test set on which network performance evaluations can be conducted. Within the training set, a validation split of $20$\% is used during training to monitor the generalisation capabilities of the network. Unlike the training loss, the loss calculated on the validation set is not used to update the model parameters during the optimisation process. 

\begin{figure}[t]
	\begin{minipage}{0.5\textwidth}
		\includegraphics[width = 1\textwidth]{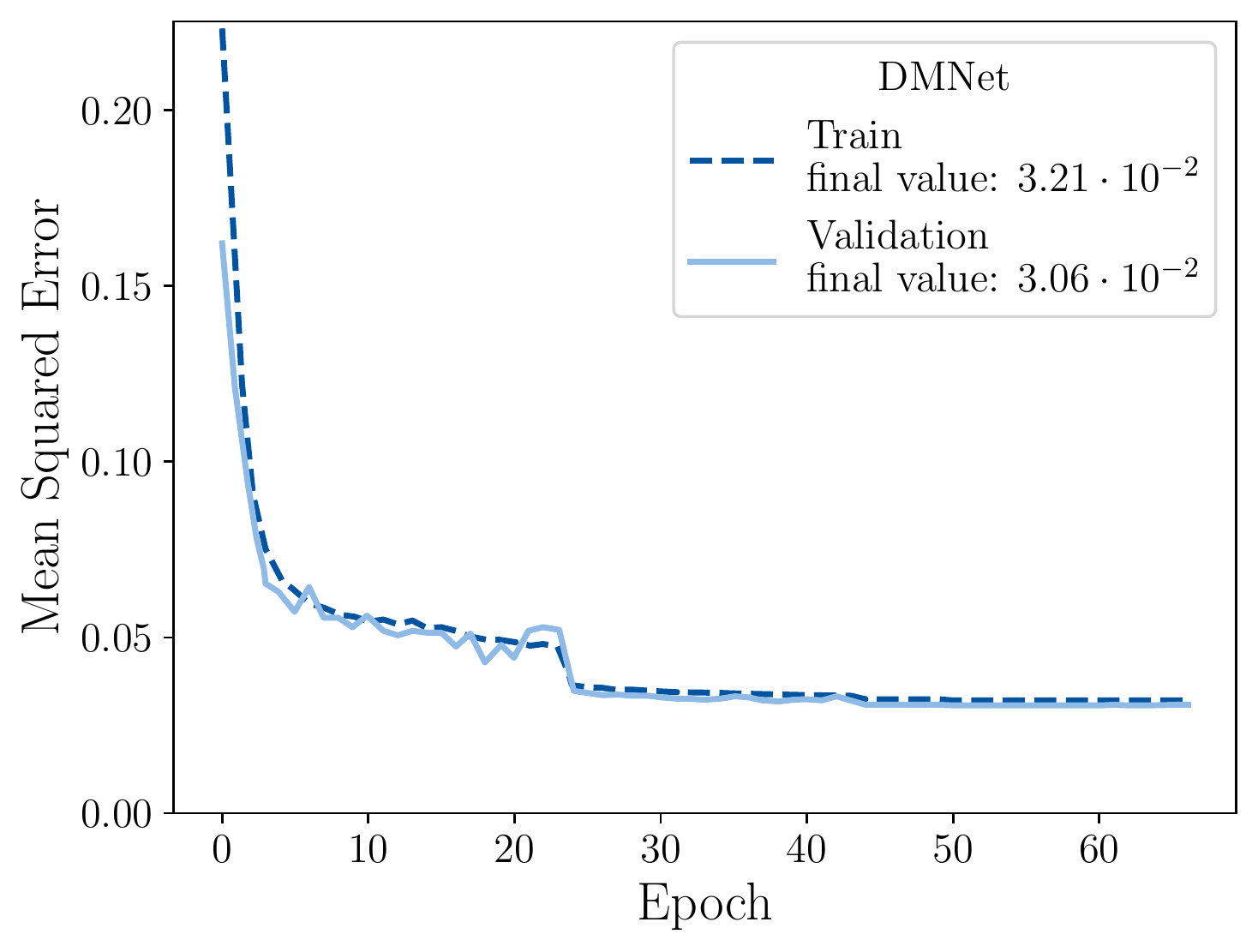}
	\end{minipage}
	\begin{minipage}{0.5\textwidth}
		\includegraphics[width = 1\textwidth]{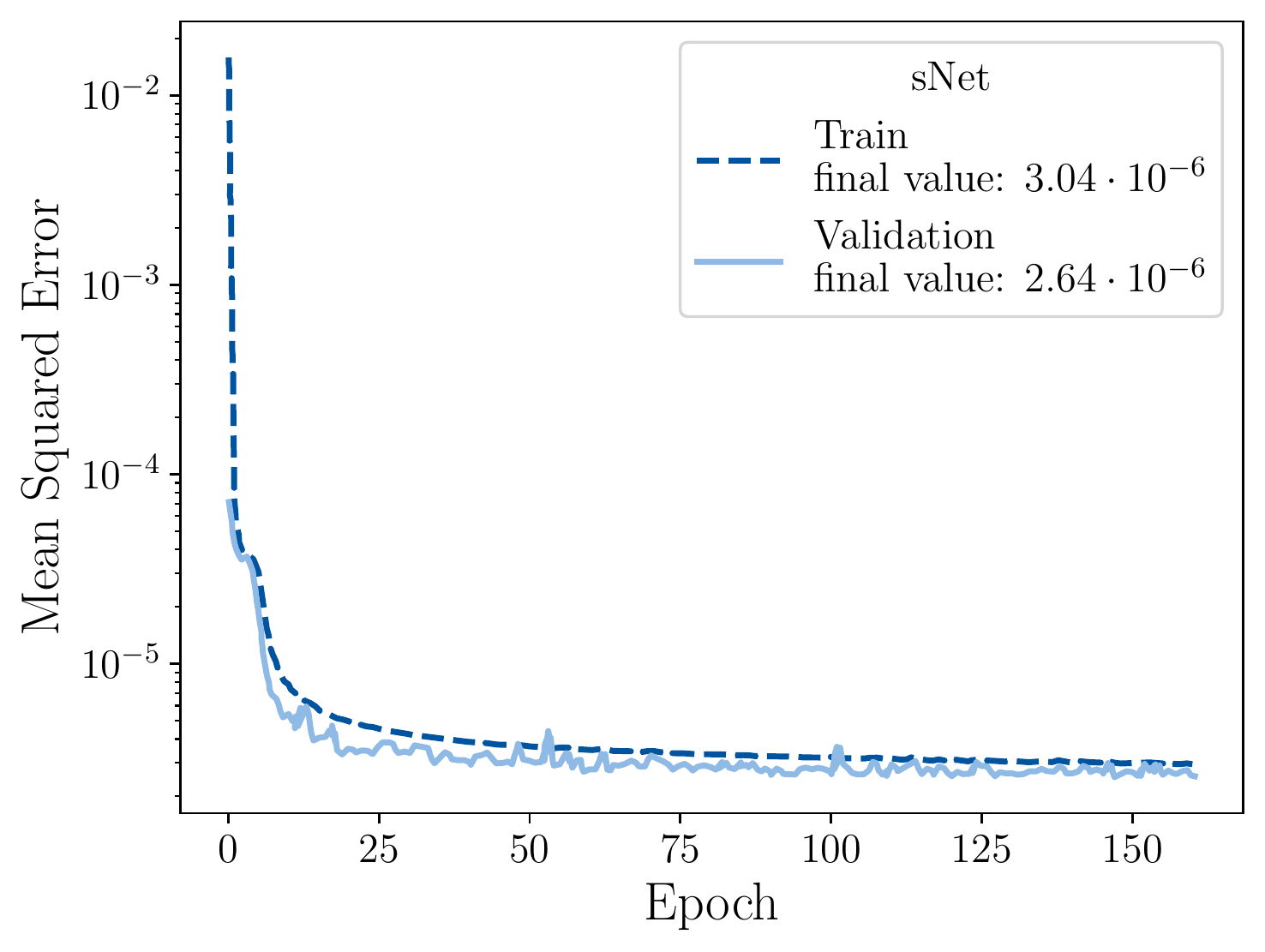}
	\end{minipage}
	\caption{Evolution of the MSE loss for the DMNet \textit{(left)} and sNet \textit{(right)} over the training epochs.}
	\label{img:loss_curves}
\end{figure}

The network training was conducted using the ADAM optimizer~\cite{adam}  and a mean squared error (MSE) loss. The initial learning rate of $10^{-2}$ is decreased during the training process, based on the behaviour of the validation loss, for an optimal convergence to a minimal loss. After the learning rate reaches its predefined minimum (lr = $10^{-5}$) the training process is terminated after 40 epochs without improvement of the validation loss, using an early stopping mechanism. This process helps ensure the convergence of the network optimisation. The MSE loss for both the training and validation loss over the training epochs is shown in figure \ref{img:loss_curves} for both ANNs. 

We performed the training on a V100-SXM2 GPU. Given the depth of the individual networks, this resulted in training durations of about $4$~minutes per epoch of the DMNet and about $12$~minutes per epoch for the sNet. 

\subsection{Validation of the Network Performance}

\begin{figure}[t]
	\centering
		\includegraphics[width = 0.8\textwidth]{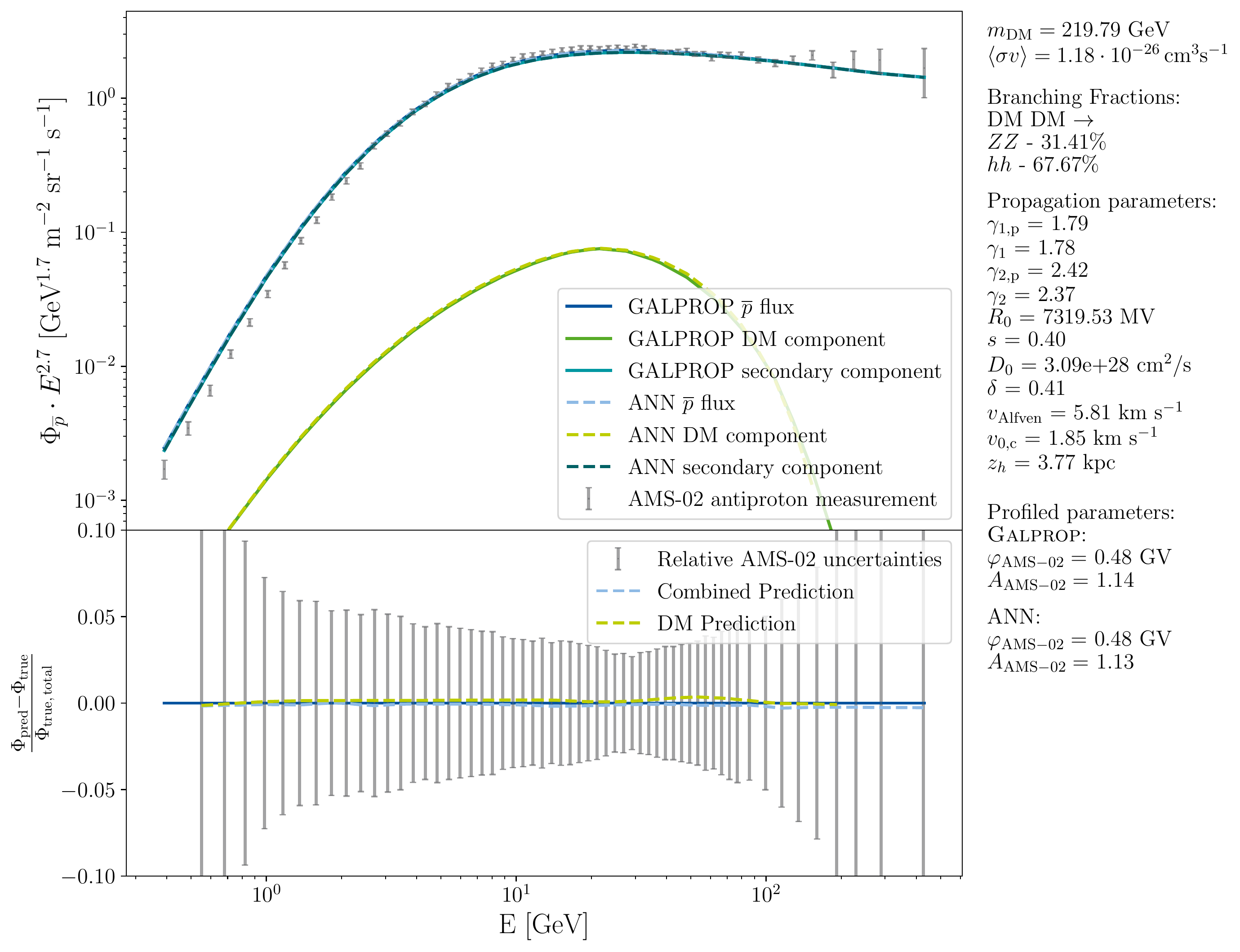}
	\caption{Exemplary comparison of the ANN versus \textsc{Galprop} antiprotons flux of only the DM component and the combination of secondary and DM component where the listed parameters and simulated fluxes are randomly sampled from the test set. Each component of the neural network flux is predicted by the individual networks. The lower panel depicts the relative difference between the \textsc{Galprop} (`true') and ANN (`predicted') fluxes with respect to the \textsc{Galprop} flux compared to the relative AMS-02 uncertainty. The listed solar modulation potential and overall normalization were inferred based on the AMS-02 data for each combined antiproton flux as described in section~\ref{sec:fitams}.}
	\label{img:example_fluxes}
\end{figure}

Training performance measures, such as the loss based on the training set, can be helpful while adjusting the architecture and hyperparameters of the networks. The usage of the networks however, requires an evaluation of their ability to replace the simulations.
Using the fully trained networks we can compare the simulated spectra from \textsc{Galprop} within the test set to the network predictions based on the same parameter point. An example for such a comparison is shown in figure~\ref{img:example_fluxes}. We show the simulated spectra and the output of the respective ANNs for both the secondary and DM component of the antiproton flux (as well as for their sum). In the top panel we depict the fluxes in physical space alongside the AMS-02 antiproton data, demonstrating that the network provides fluxes that are extremely similar to the corresponding simulations. This is illustrated even more clearly in the bottom panel, which shows the relative differences between the ANN and the simulation with respect to the simulated total antiproton flux, compared to the relative uncertainties of the AMS-02 antiproton data. Prior to plotting each CR flux we infer the solar modulation and overall normalization by maximizing the likelihood for the AMS-02 data, as outlined in section~\ref{sec:fitams}. This enables a fit to the data measured within the heliosphere and is automatically applied to each CR flux evaluated in the following. As this is not computationally expensive, it is not necessary to already include this step in the training process for the ANN. The parameters inferred for the \textsc{Galprop} and ANN fluxes respectively are in agreement with each other.
The parameter point for the specific example presented in figure~\ref{img:example_fluxes} was randomly selected from the extensive test set.

We conclude that the accuracy of the sNet is fully sufficient: the relative difference between the fluxes predicted by the ANN and the simulated \textsc{Galprop} fluxes are always well below the relative uncertainty of the AMS-02 measurements. The architecture and training process used for the sNet can analogously be applied to train an ANN on proton and Helium spectra based on the same \textsc{Galprop} simulation set, achieving a comparable accuracy. We provide additional details on these networks in appendix~\ref{app:p_and_He}.

Given that the DMNet is trained on a parameter space of much higher dimensionality, it is unsurprising that its predictions are on average less accurate than the ones of the sNet. Indeed when calculating the relative differences between simulations and network predictions for the DM component only, we find that only 72\% of samples lie on average within the AMS-02 relative uncertainties. However, it is essential to realise that in any realistic setting the DM component will only constitute a subdominant contribution to the antiproton flux. Indeed, if the DM contribution in a given bin significantly exceeds the uncertainty of the AMS-02 data (which is typically at the level of 5\%) the model is expected to be excluded. 

In order to provide more realistic estimates of the general accuracy and stability of the DMNet performance within the test set, we therefore focus on DM signals that contribute 5\% to the total antiproton flux in the bin where the relative DM contribution is largest. We then calculate the differences between simulations and network output relative to the total antiproton flux. This approach shows that even if the DMNet itself is only accurate at the level of 10\%, the total antiproton flux can still be predicted with an accuracy at the sub-percent level for allowed DM models.

\begin{figure}[t]
	\centering
		\includegraphics[width = 1\textwidth]{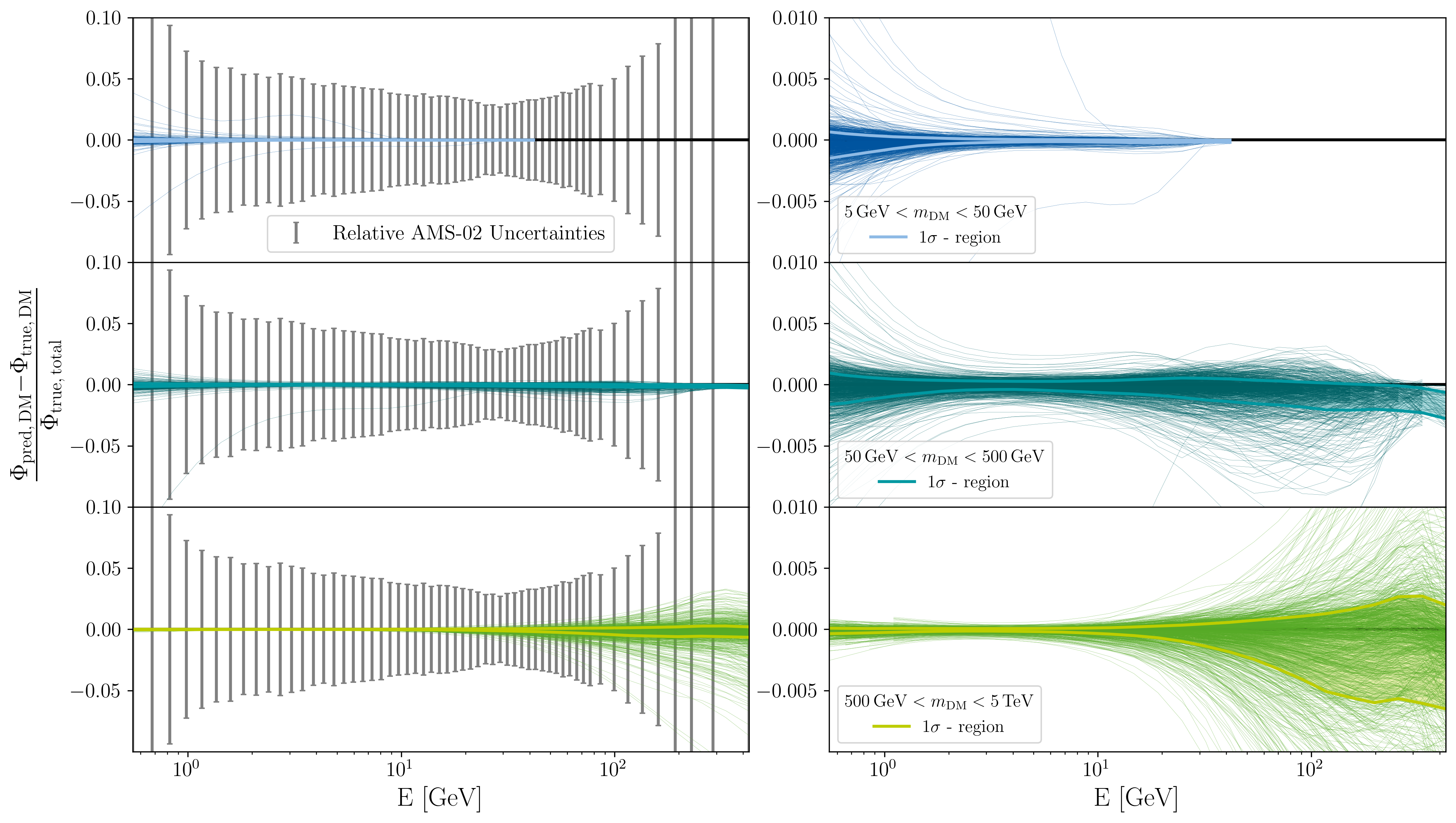}
	\caption{Relative deviations between predicted and simulated fluxes for the DM component binned into three mass bins and their 68 percentile. Each of the panel shows 1000 samples from the test set. As in the lower panels in figure~\ref{img:example_fluxes}, in the left panel we again compare this to the benchmark of the relative AMS-02 uncertainties. The right panel shows a zoomed-in version of the left panel in the interval [-0.010, 0.010].}
	\label{img:residual_bands}
\end{figure}

In figure~\ref{img:residual_bands}, we show this accuracy estimate for a total of 3000 DM component samples from the test set (1000 samples each for three different mass bins corresponding to the three different rows). Here we compute the deviation between DMNet prediction and \textsc{Galprop} simulation to the corresponding total antiproton flux, as in the lower panel of figure~\ref{img:example_fluxes}. Since the deviations are found to be miniscule when compared to the AMS-02 relative uncertainties, we provide in the right column of figure~\ref{img:residual_bands} a zoomed-in version. It can be seen that the uncertainty bands (containing the central 68\% of the network predictions) are typically at the level of 0.1\% and do not exhibit any systematic shifts nor any significant dependence on the DM mass.  

In the following we will be interested in comparing the total antiproton flux to data in order to determine which DM signals are allowed by observations. The comparison between the network accuracy and the AMS-02 uncertainties in figure~\ref{img:residual_bands} clearly shows that it is fully sufficient for this purpose to use the ANNs instead of running \textsc{Galprop}. Indeed, we will show explicitly in the next section that both approaches lead to very similar values for the $\chi^2$ statistic described in section~\ref{sec:fitams}.

The fully trained networks, as described in this section and appendix~\ref{app:p_and_He}, are publicly available as \textsc{DarkRayNet} at \url{https://github.com/kathrinnp/DarkRayNet}. In this repository, we provide an interface to easily access flux predictions for the corresponding CR species. This tool can for example be used for indirect DM searches as we outline in our analysis in the subsequent section.  

\section{Constraining the dark matter annihilation cross section}
\label{sec:constraints}

\subsection{Statistical method}
\label{sec:marg_importance}

The ANNs described in the previous section enable us to obtain predictions of the primary and secondary antiproton flux as a function of the DM parameters $\xDM$ and the propagation parameters $\tprop$. Given data from observations we can then construct a likelihood function $\mathcal{L}(\xDM,\tprop)$. We emphasise that, given suitable predictions for the CR fluxes, this likelihood is quick to evaluate and therefore does not need to be predicted by the ANN. This has the significant advantage that the ANN does not need to learn the various fluctuations that may be present in the data.

The likelihood function can then be used to constrain both $\xDM$ and $\tprop$. In the present work we primarily focus on the constraints on the DM parameter space, meaning that we will treat the propagation parameters simply as nuisance parameters that need to be varied in order to draw robust conclusions. The two main ways to achieve this is to either calculate the profile likelihood
\begin{equation}
 \hat{\mathcal{L}}(\xDM) = \mathcal{L}(\xDM,\hat{\bm{\theta}}_\text{prop}(\xDM)) \; ,
\end{equation}
where $\hat{\bm{\theta}}_\text{prop}(\xDM)$ denote the propagation parameters that maximise the likelihood for given DM parameters $\xDM$, or to calculate the marginalised likelihood
\begin{equation}
 \bar{\mathcal{L}}(\xDM) = \int \mathcal{L}(\xDM,\tprop) p(\tprop) \mathrm{d}\tprop \; ,
\end{equation}
where $p(\tprop)$ denotes the prior probability for the propagation parameters. Given sufficiently constraining data, the profile likelihood and the marginalised likelihood are expected to be similar and the dependence of the result on the chosen priors should be small. We find that this is largely true for the case considered here, with some notable exceptions to be discussed below.

From the point of view of our machine learning approach, however, the two ways of varying the nuisance parameters are very different. The profile likelihood depends only on the antiproton flux for a single value of $\tprop$, meaning that highly accurate predictions are needed close to the maximum of the likelihood. For extreme choices of the DM parameters this maximum may be pushed to corners of parameter space where the network has not been sufficiently trained. A single outlier in the prediction will then completely bias the result and lead to numerical instabilities {when sampling the parameter space}. This makes accurate calculations of the profile likelihood a highly challenging task.

The marginalised likelihood, on the other hand, depends on the likelihood across a range of propagation parameters, which should have substantial overlap with the parameter regions seen during training. The impact of individual outliers in the predictions is also reduced significantly compared to the case of the profile likelihood, making the calculation of marginalised likelihoods based on ANN predictions more robust. Nevertheless, the challenge remains to ensure that results are not biased by regions of parameter space where only little training has been performed. In the present work we address this challenge using the technique of importance sampling~\cite{mcbook}, which we describe in the following.\footnote{For a different approach to Bayesian analyses of cosmic ray propagation with the help of neural networks we refer to Ref.~\cite{Johannesson:2016rlh}.}

First of all, we note that an approximate marginalisation can be performed by drawing a random sample of parameter points $\ti$ from the prior probability $p(\tprop)$ and calculating the sum
\begin{equation}
 \bar{\mathcal{L}}(\xDM) \approx \frac{1}{N} \sum_{i=1}^N \mathcal{L}(\xDM, \ti) \; .
\end{equation}
In fact, the same can be done by drawing a random sample from any probability distribution function $q(\tprop)$ provided the individual points are reweighted accordingly (so-called importance sampling):
\begin{equation}
 \bar{\mathcal{L}}(\xDM) \approx \frac{\sum_{i=1}^N \mathcal{L}(\xDM, \ti) \frac{p(\ti)}{q(\ti)}}{\sum_{i=1}^N \frac{p(\ti)}{q(\ti)}} \; .
 \label{eq:sample}
\end{equation}
A particularly interesting case is that $q(\tprop)$ is taken to be the posterior probability for the propagation parameters in the absence of a DM signal, i.e.
\begin{equation}
	q(\tprop) \propto \mathcal{L}(\xDM = 0, \tprop) \, p(\tprop) \equiv \mathcal{L}_0(\tprop) \, p(\tprop) \, .
\end{equation}
In this case $p(\ti) / q(\ti) \propto 1/\mathcal{L}_0(\theta_i)$ and hence
\begin{equation}
 \bar{\mathcal{L}}(\xDM) \approx \frac{\sum_{i=1}^N  \frac{\mathcal{L}(\xDM, \ti)}{\mathcal{L}_0(\ti)}}{\sum_{i=1}^N \frac{1}{\mathcal{L}_0(\ti)}} \; .
\end{equation}

The great advantage of this approach is that the likelihood is only evaluated for plausible values of the propagation parameters, meaning for values that give a large posterior probability in the absence of a DM signal. These are exactly the same parameter regions on which we have focused for the training of the ANNs described above. Indeed, it is possible to generate the training data and sample the posterior probability using the same prior probabilities and the same \textsc{MultiNest} runs such that a large overlap between the two is ensured.\footnote{We emphasise that the posterior sample is not part of the training data, i.e.\ the ANN is never evaluated on the exact same values seen during training.} Another significant advantage is that it is straight-forward to include additional constraints on the propagation parameters that are independent of the DM parameters and therefore not part of the ANN training. For example, to also include likelihoods for proton data $\mathcal{L}_p$ and He data $\mathcal{L}_\text{He}$, it is sufficient to draw a sample from the joint posterior
\begin{equation}
 q(\tprop) \propto \mathcal{L}_0(\tprop) \, \mathcal{L}_p(\tprop) \, \mathcal{L}_\text{He}(\tprop) \, p(\tprop) \, .
\end{equation}

To conclude this discussion, we note that in the case that the likelihood can be written in terms of a $\chi^2$ function, $\mathcal{L} \propto e^{- \chi^2/2}$, we can define a marginalised $\chi^2$ function as $\bar{\chi}^2(\xDM) \equiv -2 \log \bar{\mathcal{L}}(\xDM)$. Importance sampling then yields
\begin{equation}
	\bar{\chi}^2(\xDM) = - 2 \log {  \frac{\sum_{i=1}^N  \exp{ \left(- \frac{\Delta\chi^2(\xDM, \tprop)}{2} \right)}} {\sum_{i=1}^N  \exp{ \left(\frac{\chi^2_0(\tprop)}{2} \right)} }} \, ,
	\label{eq:marg_chi}
\end{equation}
where $\chi^2_0(\tprop) = \chi^2(\xDM = 0, \tprop)$ and $\Delta\chi^2(\xDM, \tprop) = \chi^2(\xDM,\tprop) - \chi^2_0(\tprop)$.
To calculate confidence intervals and exclusion limits for the DM parameters, we then define
\begin{equation}
	\Delta \bar{\chi}^2(\xDM) = \bar{\chi}^2(\xDM) - \bar{\chi}^2_0\, .
	\label{eq:upper_bound}
\end{equation}
Hence, $\Delta \bar{\chi}^2 < 0$ corresponds to a preference for a DM signal, while parameter points with $\Delta \bar{\chi}^2 > 3.84$ can be excluded at 95\% confidence level.\footnote{Note that although our treatment of nuisance parameters is motivated by Bayesian statistics, we still interpret the resulting marginalised likelihood using frequentist methods, such that there is no need to choose priors for the DM parameters.}

\subsection{Example A: Single Dark Matter Annihilation Channel}
\label{sec:example_A}

\begin{figure}[t]
	\begin{minipage}{0.5\textwidth}
		\includegraphics[width = 1\textwidth]{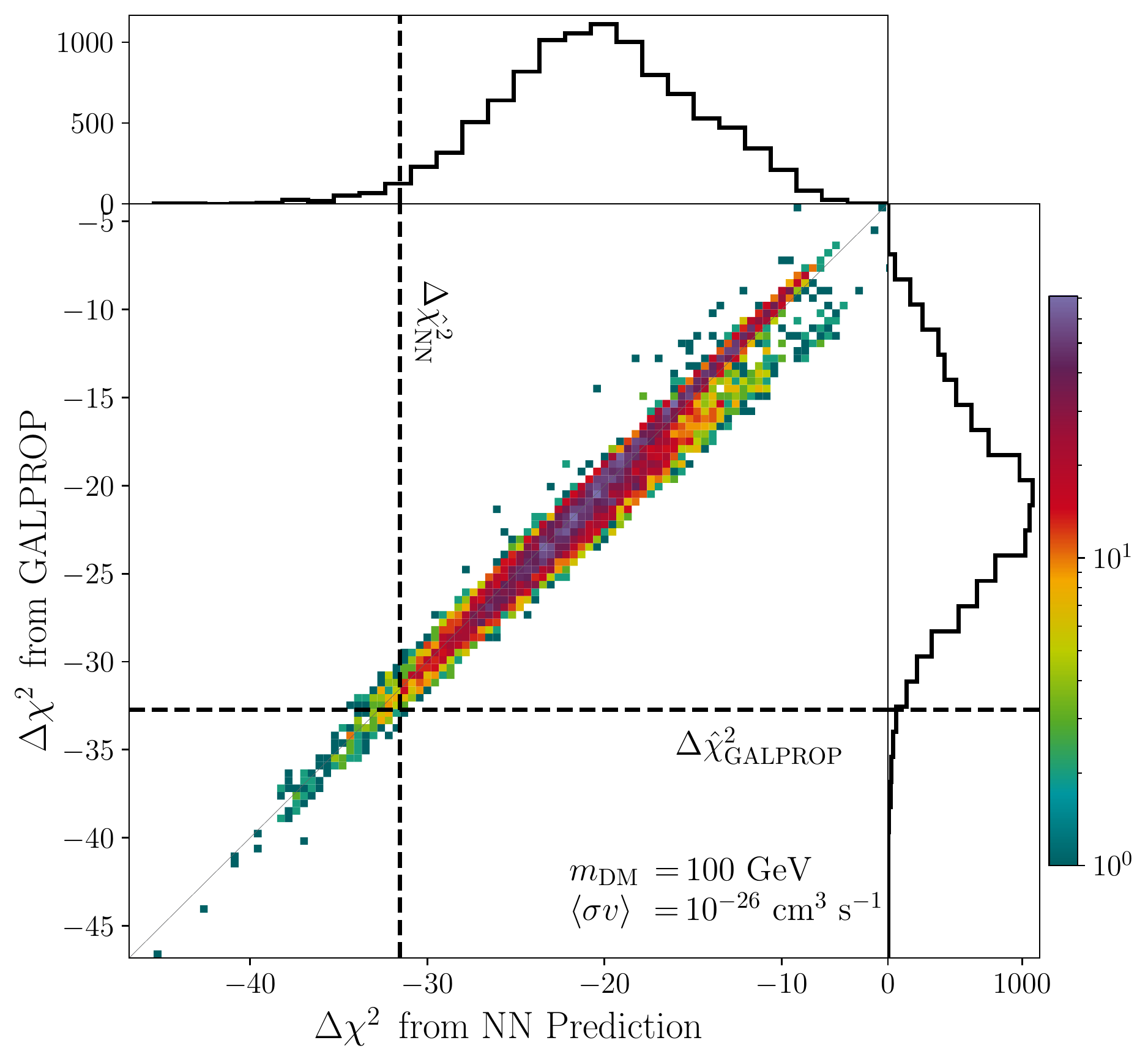}
	\end{minipage}
	\begin{minipage}{0.5\textwidth}
		\includegraphics[width = 1\textwidth]{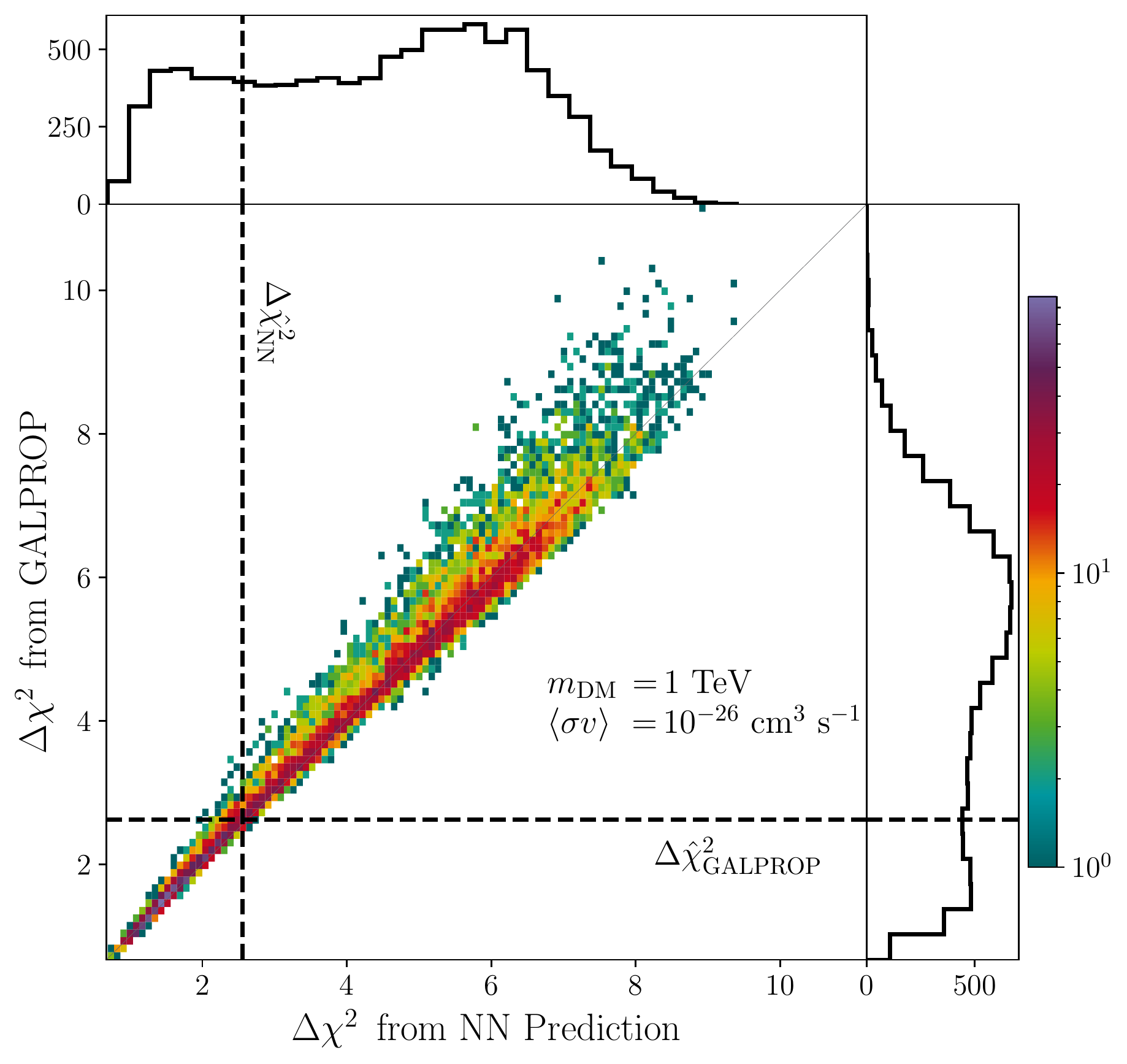}
	\end{minipage}
	\caption{One and two dimensional histograms of $\Delta \chi^2$ for the AMS-02 antiproton measurement based on the antiproton fluxes provided by the Neural Network and \textsc{Galprop} for different combinations of propagation parameters. 
	We consider the annihilations of DM particles with $m_\text{DM} = 100$ GeV \textit{(left)} and 1 TeV \textit{(right)} into $b \overline{b}$ with a cross section of $\langle \sigma v \rangle = 10^{-26}$ cm$^3$ s$^{-1}$. The values for $\Delta \hat{\chi}^2$ indicated by the black dashed lines represent the marginalised values obtained by the importance sampling technique described in section~\ref{sec:marg_importance}. } 
	\label{img:chi_comp}
\end{figure}

Let us first consider a frequently-used benchmark scenario and assume that the DM particles annihilate exclusively into pairs of bottom quarks, such that the injection spectrum is fully characterised by the (velocity-independent) annihilation cross section $\langle \sigma v \rangle$ and the DM mass $m_\text{DM}$. As a first step, we can then calculate $\Delta \chi^2(m_\text{DM}, \langle \sigma v \rangle, \tprop)$ for different values of the propagation parameters. Figure~\ref{img:chi_comp} compares the results that we obtain when using the ANN predictions of the antiproton flux and when employing \textsc{Galprop}. The two panels correspond to different values of the DM mass and use the same 10122 sets of propagation parameters drawn randomly from the posterior distribution $q(\tprop)$ as discussed above. In both cases we find a very strong correlation between the two ways of calculating $\Delta \chi^2$ ($r > 0.98$). Indeed, for 95\% of parameter points the absolute difference in $\Delta \chi^2$ is smaller than $2.1$ ($0.9$) for $m_\text{DM} = 100\,\mathrm{GeV}$ ($m_\text{DM} = 1\,\mathrm{TeV}$), confirming the excellent performance of our ANN.

In each case we use a dashed line to indicate $\Delta \bar{\chi}^2$ as defined in eq.~(\ref{eq:marg_chi}). We emphasise that, since we average over $\exp(-\Delta \chi^2 / 2)$, the final result is dominated by the points with the smallest $\Delta \chi^2$. Again, we find very good agreement between the marginalised $\Delta \chi^2$ obtained from the ANN and from \textsc{Galprop}. The values obtained in the left panel correspond to a substantial preference for a DM signal, while the parameter point considered in the right panel is slightly disfavoured by data. Although the value $\Delta \bar{\chi}^2 = -31.5$ ($-32.7$) that we obtain for $m_\text{DM} = 100 \, \mathrm {GeV}$ from the ANN (\textsc{Galprop}) would at face value correspond to quite a significant excess, we {emphasize that our set-up is not designed to provide an accurate characterisation of this excess. In particular we} caution the reader that due to our simplified implementation of AMS-02 data (in particular neglecting correlations) this number should be interpreted with care. We expect that a more detailed analysis of AMS-02 data would lead to a much lower significance.

Comparing the evaluations of the marginalised $\Delta \chi^2$ with the ANN and \textsc{Galprop} respectively, the reduction of the computational cost achieved with our neural network method becomes apparent. For the ANN the prediction of the set of CR fluxes for each of the specific DM parameter points only takes $\mathcal{O}(1)$ cpu second in total for the 10122 parameter points, but the calculation of the respective $\chi^2$ while inferring the solar modulation potential takes up the majority of the computation time ($\mathcal{O}(10)$ cpu seconds in total). This time is however negligible compared to the \textsc{Galprop} simulations which take $\mathcal{O}(10)$ cpu hours to obtain the same number of CR fluxes.

\begin{figure}[t]
	\begin{minipage}{0.5\textwidth}
		\includegraphics[width = 1\textwidth]{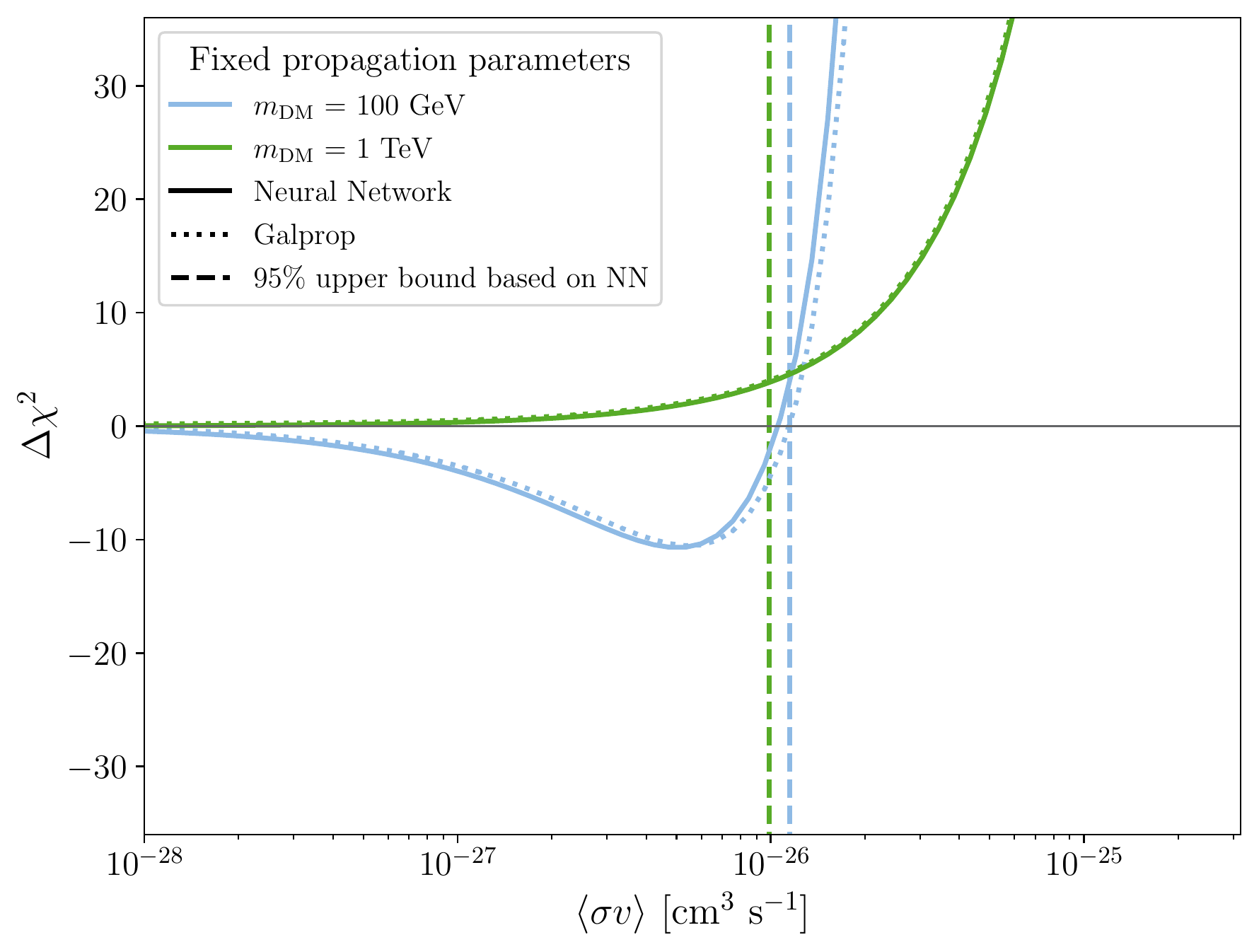}
	\end{minipage}
	\begin{minipage}{0.5\textwidth}
		\includegraphics[width = 1\textwidth]{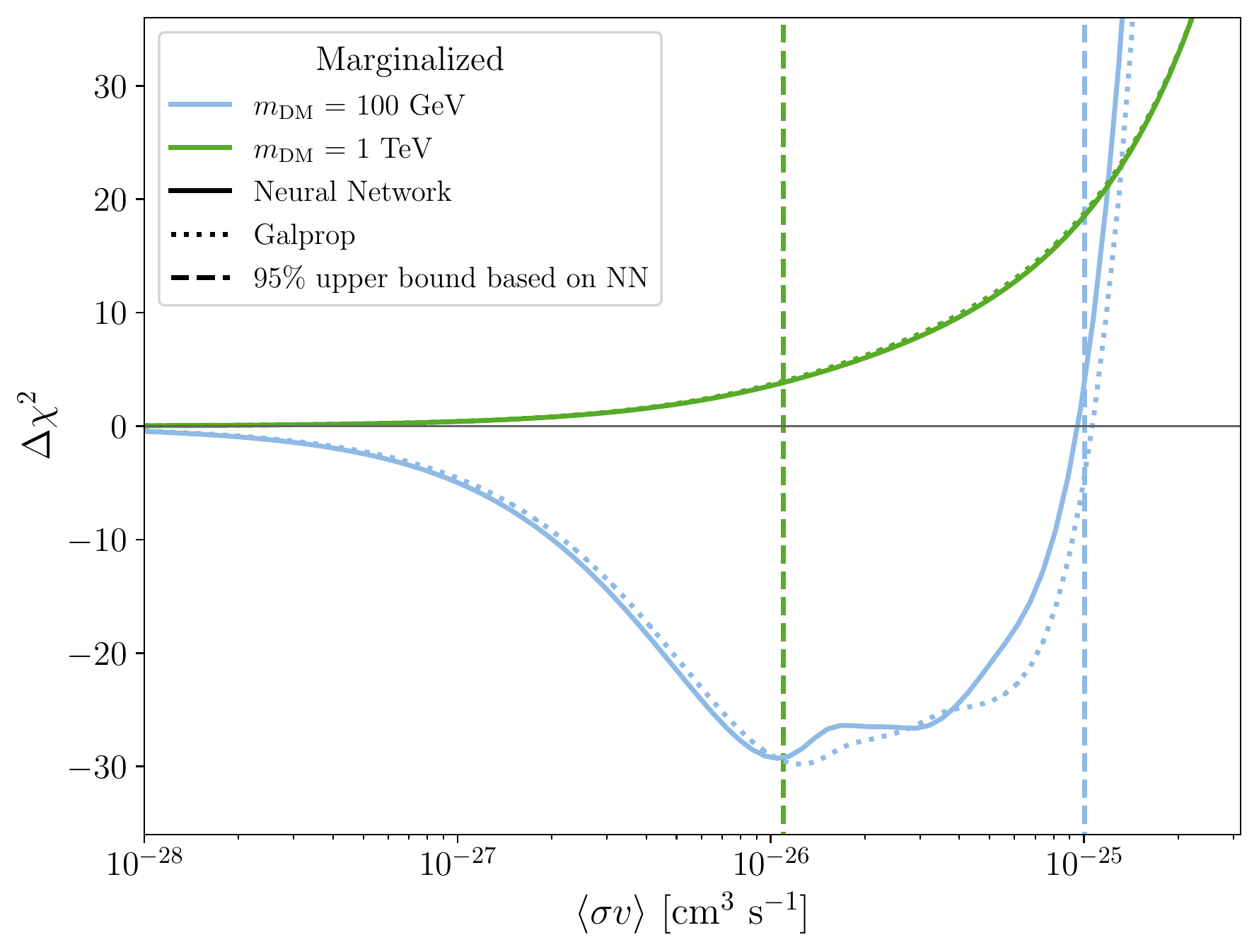}
	\end{minipage}
	\caption{$\Delta \chi^2$ for the AMS-02 antiproton measurement based on the antiproton fluxes provided by the Neural Network and \textsc{Galprop} as a function of $\langle \sigma v \rangle$ and for different values of $m_\text{DM}$. We assume a dominant DM DM $\rightarrow \, b \overline{b}$ annihilation in each case. \textit{Left:} Propagation parameters are fixed to the best-fit values in a frequentist setup when only secondary antiprotons are considered (see table~\ref{tab:param_ranges_v2}). \textit{Right:} Propagation parameters are marginalised over using importance sampling. We also include the 95 \% upper bound values of the annihilation cross section following eq.~(\ref{eq:upper_bound}).}
	\label{img:chi_dist_zoom}
\end{figure}

A complementary perspective to the results in figure~\ref{img:chi_comp} is provided in figure~\ref{img:chi_dist_zoom}, which shows $\Delta \chi^2$ as a function of $\langle \sigma v \rangle$ for different values of the DM mass. In the left panel we fix the propagation parameters to their best-fit values in the absence of a DM signal (see table~\ref{tab:param_ranges_v2}), while in the right panel we marginalise over all propagation parameters using importance sampling. Solid (dotted) curves correspond to the ANN (\textsc{Galprop}) predictions and again show excellent agreement. The horizontal dashed lines indicate the 95\% confidence level upper bound on $\langle \sigma v \rangle$ obtained following eq.~(\ref{eq:upper_bound}).

As expected, allowing variations in the propagation parameters generally leads to smaller values of $\Delta \chi^2$ and hence relaxes the upper bounds on the annihilation cross section. This effect is most dramatic for the case $m_\text{DM} = 100 \, \mathrm{GeV}$ (blue line), where there is a preference for a DM signal in the data and hence the exclusion limit is relaxed by about an order of magnitude. The small bumps in the blue curve in the right panel are a result of the finite size of the sample of propagation parameters used for the marginalisation and result from the approximation made in eq.~(\ref{eq:sample}).

\begin{figure}[t]
		\includegraphics[width = 1\textwidth]{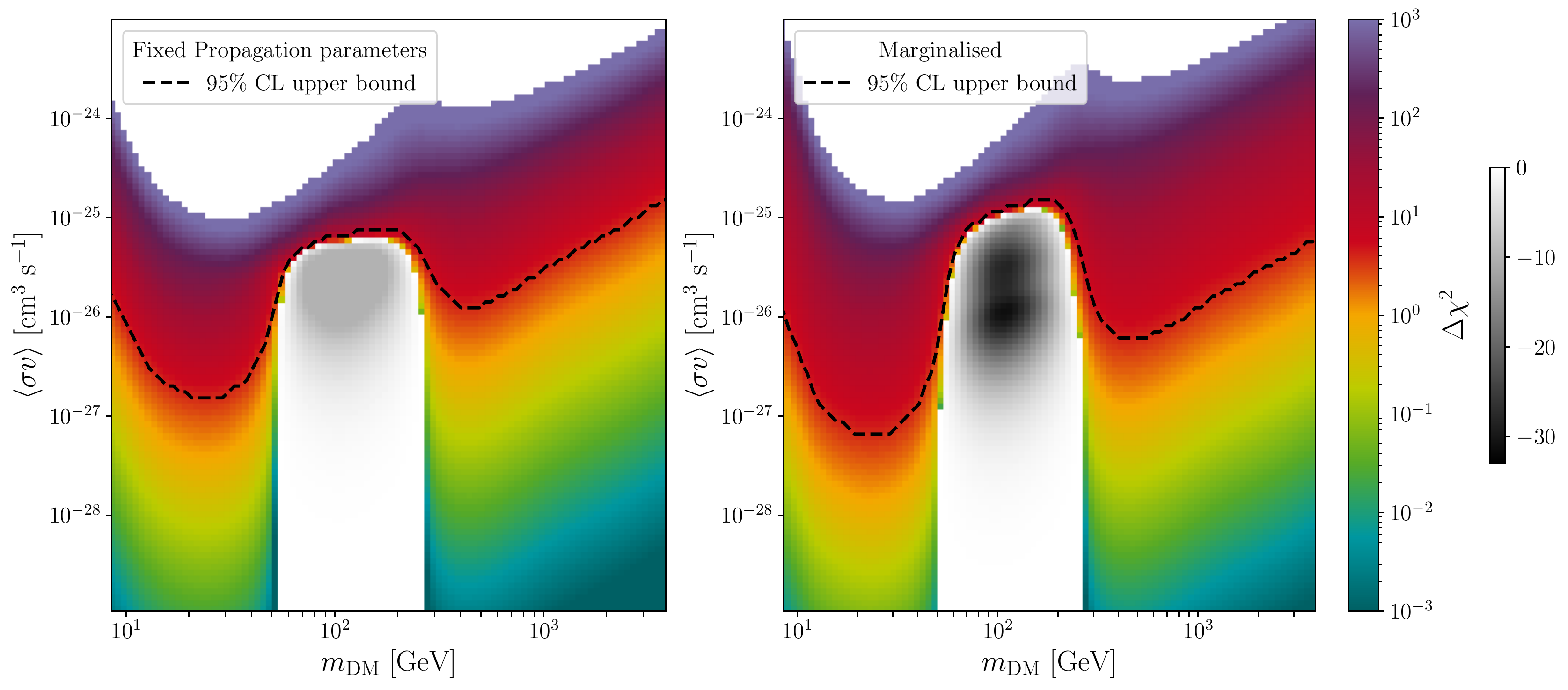}
	\caption{$\Delta \chi^2$ for the AMS-02 antiproton measurement as a function of $\langle \sigma v \rangle$ and $m_\text{DM}$ using {the fixed propagation parameters specified in table~\ref{tab:param_ranges_v2}} (\emph{left}) and performing the marginalisation via importance sampling (\emph{right}). The dashed lines represent the 95~\% CL upper bounds on the annihilation cross section. The white regions in the upper part of each panel correspond to $\Delta \chi^2 > 1000$ and are excluded to improve numerical stability.}
	\label{img:bounds_bb}
\end{figure}

Repeating this procedure for different values of the DM mass, we can obtain exclusion limits on $\langle \sigma v \rangle$ as a function of $m_\text{DM}$. These are shown in figure~\ref{img:bounds_bb} for the case of fixed propagation parameters (left) and when marginalising over propagation parameters (right). The colour shading indicates parameter regions where $\Delta \chi^2 > 0$, such that a DM signal is disfavoured, while greyscale is used to indicate parameter regions where $\Delta \chi^2 < 0$ such that a DM signal is preferred. We find that this is the case for DM masses in the range $50\text{--}250\,\mathrm{GeV}$. Again, marginalisation leads to relaxed exclusion bounds  and an increased preference for a DM signal. We reiterate however that the magnitude of this preference is likely overestimated in our analysis.

To assess the impact of marginalisation let us finally compare our results with those obtained using a profile likelihood. As discussed in section~\ref{sec:marg_importance}, special care needs to be taken when using the ANN predictions to calculate a profile likelihood in order to ensure that the result is not dominated by regions of parameter space with insufficient training data. We achieve this goal by restricting the allowed parameter regions as follows: $0.1 < s < 0.6$, $1 \, \mathrm{GV} < R_0 < 10 \, \mathrm{GV}$, $0.35 < \delta < 0.6$ and $2.3 < \gamma_{2,(p)} < 2.5$. We then use \textsc{MultiNest} to explore the remaining parameter space for fixed values of the DM mass and varying annihilation cross section in order to find the largest value of $\langle \sigma v \rangle$ such that $\Delta \hat{\chi}^2(m_\text{DM}, \langle \sigma v \rangle) \equiv  -2 \Delta \log \hat{\mathcal{L}}((m_\text{DM}, \langle \sigma v \rangle)) < 3.84$. Repeating this procedure for different values of $m_\text{DM}$ then yields the exclusion limit.

\begin{figure}[t]
	\centering
	\includegraphics[width = 0.7\textwidth]{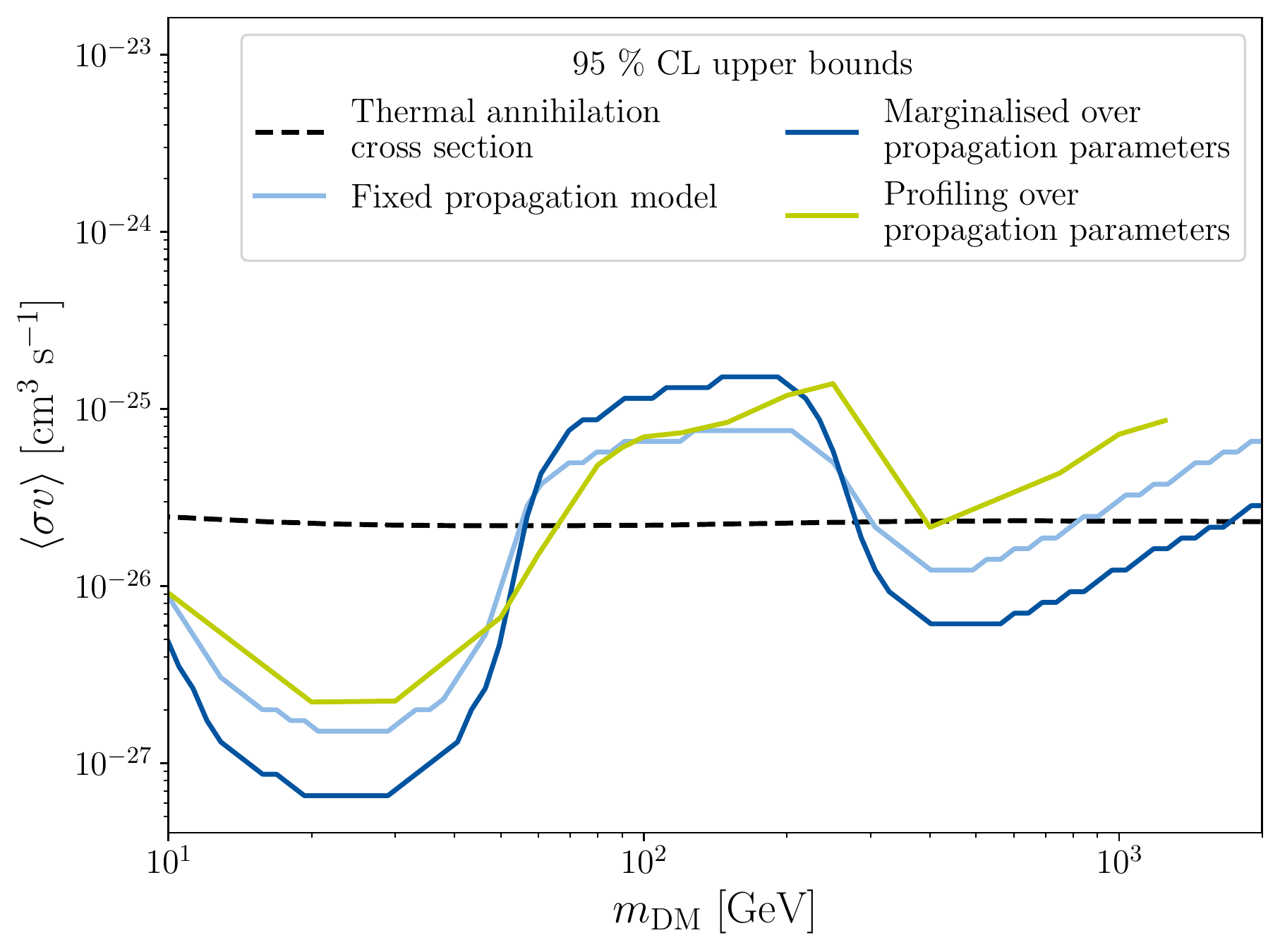}
	\caption{A comparison of the 95~\% CL exclusion bounds in figure~\ref{img:bounds_bb} (blue and light blue) with the bounds obtained when profiling over the propagation parameters using the CR spectra provided by our ANNs (green). The black dashed line indicates the thermal annihilation cross section for WIMPs from \cite{Steigman2012}.}
	\label{img:bounds_bb_compare}
\end{figure}

The results are shown in figure~\ref{img:bounds_bb_compare} together with the exclusion limits obtained for fixed propagation parameters and when marginalising over propagation parameters as shown in figure~\ref{img:bounds_bb}. We find that in most regions of parameter space the profile likelihood approach yields somewhat weaker exclusion limits than the marginalisation. Such a difference is to be expected whenever substantial tuning in the propagation parameters is required in order to accommodate a DM signal. For example, for $m_\text{DM} = 1 \, \mathrm{TeV}$ and $\langle \sigma v \rangle = 5 \times 10^{-26} \, \mathrm{cm^3 \, s^{-1}}$ we find that $\Delta \hat{\chi}^2 < 3.84$ can be achieved only if $D_0$, $v_{0,c}$ and $z_\mathrm{h}$ all take values close to their lower bounds. Such a tuning is not penalised in the profile likelihood, but the contribution of these solutions to the marginalised likelihood will be suppressed according to the small volume of the viable parameter space. The same conclusion can be reached from the right panel of figure~\ref{img:chi_comp}: Although there are sets of propagation parameters that yield $\Delta \chi^2 \approx 0$, most parameter combinations give significantly larger $\Delta \chi^2$, such that marginalisation leads to $\Delta \hat{\chi}^2 \approx 2.6$, close to the 95\% confidence level upper bound. In other words, the difference between the two approaches is a direct consequence of the different statistical methods and not an artefact of the ANN predictions.

In general the dependence of the DM limit on the chosen value for the halo height is very well known. To first order the normalisation of the 
DM flux is proportional to $z_\mathrm{h}$ and thus the DM limit is anti-proportional to $z_\mathrm{h}$ as again nicely demonstrated in 
a very recent analysis \cite{Genolini:2021doh}.
The CR fit conducted in section \ref{sec:cr} varies $z_\mathrm{h}$ between 2 and 7 kpc. Because of the well-known $z_\mathrm{h}$-$D_0$ 
degeneracy the resulting posterior of $z_\mathrm{h}$ is almost flat in the entire fit range. 
The DM limit derived from the marginalisation of the $\Delta \hat{\chi}^2$ should be understood to refer to 4.8 kpc, 
namely the average value of $z_\mathrm{h}$ in the posterior. This is in perfect agreement with recent analyses 
of secondary fluxes by AMS-02~\cite{Evoli:2019iih,Weinrich:2020ftb,Luque:2021joz,Korsmeier:2021brc}.
On the other hand, when limits are derived in a frequentist approach and in the absence of a DM preference, 
$z_\mathrm{h}$ values are pushed towards the lower boundary of the fit range at 2 kpc. 
This again explains the difference between the marginalised and profiled limit in the figure~\ref{img:bounds_bb_compare}.
One possible way to study the $z_\mathrm{h}$ dependence explicitly in the marginalisation framework is to 
further restrict the range of $z_\mathrm{h}$. 

The differences between marginalised and profiled limits are particularly relevant given how they affect the conclusions drawn from figure~\ref{img:bounds_bb_compare}. When using the marginalised likelihood we find that the thermal cross section (indicated by the black dashed line) can be excluded for DM masses in the range $300\text{--}2000\,\mathrm{GeV}$, implying that WIMP models in this mass range can only be viable if the injection of antiprotons are suppressed. When using the profile likelihood, on the other hand, almost the entire mass range above $70 \, \mathrm{GeV}$ is found to be viable. 	We note that the agreement between the frequentist and Bayesian approach will improve with a better determination of $z_\mathrm{h}$ as expected from the analysis of the forthcoming Be isotope measurements by AMS-02 \cite{Derome:ICRC2021}.

In addition to the reduction in computing time achieved when using the ANN instead of \textsc{Galprop}, we find that the use of importance sampling leads to another improvement compared to the more conventional profiling approach.
Crucially, our marginalisation using importance sampling is based on a fixed set of 10122 data points in the propagation model, which can be evaluated in parallel. The ANN therefore gives a negligible contribution to the time needed to calculate the upper bound on the annihilation cross section for each of the 100 mass bins shown in figures~\ref{img:bounds_bb} and \ref{img:bounds_bb_compare}. 
For the profiling approach on the other hand the evaluation of the data points cannot be performed in parallel by the ANN due to their sampling. This leads to an increase in computation time, such that  the speed-up of the runtime when using the ANN instead of \textsc{Galprop} is reduced to two orders of magnitude (rather than three orders of magnitude for importance sampling).

\subsection{Example B: Scalar Singlet Dark Matter}

We now illustrate the use of the ANN for the analysis of a specific model of DM with a singlet scalar field $S$.  Imposing a $Z_2$ symmetry, $S \to -S$, the scalar particle is stable and thus a DM candidate. The Lagrangian of this scalar singlet DM (SSDM) model reads~\cite{Silveira:1985rk,McDonald:1993ex,Burgess:2000yq} 
\begin{equation}
{\cal L} = {\cal L}_\text{SM} + \frac 12 \partial_\mu  S  \partial^\mu  S  - \frac12 m_{S,0}^2S^2- \frac 14 \lambda_S  S^4- \frac 12 \lambda_{H\!S}\, S^2 H^\dagger H\,, 
\label{eq:lagr}
\end{equation}
where ${\cal L}_\text{SM}$ is the Standard Model Lagrangian and $H$ is the Standard Model Higgs field. After electroweak symmetry breaking, the last three terms of the Lagrangian become
\begin{equation}
{\cal L} \supset  - \frac12 m_{S}^2\, S^2- \frac 14 \lambda_S\,  S^4 - \frac 14 \lambda_{H\!S}\, h^2 S^2 - \frac {1}{2} \lambda_{H\!S}\, v h S^2\,,
\label{eq:ewbr}
\end{equation}
with $H = (h+v, 0)/\sqrt{2}\,$, $v = 246\,$GeV, and where we introduced the physical mass of the singlet 
field, $m_S^2  = m_{S,0}^2 + \lambda_{H\!S} \,v^2 / 2$. The DM phenomenology of the SSDM has been extensively studied in the literature, see e.g.\ \cite{Cline:2013gha,Beniwal:2015sdl,Cuoco:2016jqt,Cuoco:2017rxb,GAMBIT:2017gge,Athron:2018ipf} and references therein. 

\begin{figure}[t]
	\begin{minipage}{0.435\textwidth}
		\includegraphics[width = 1\textwidth]{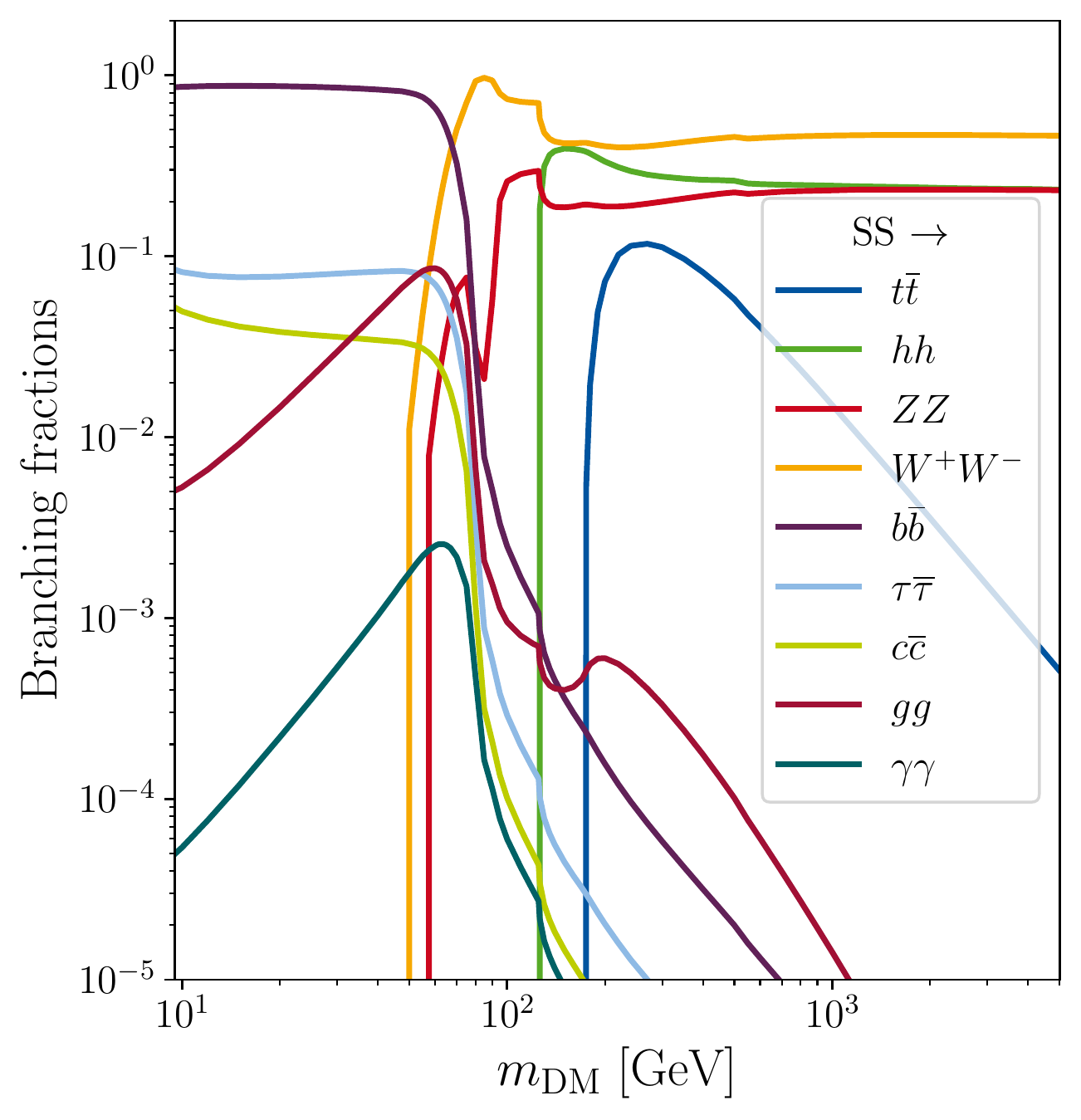}
	\end{minipage}
	\begin{minipage}{0.565\textwidth}
		\includegraphics[width = 1\textwidth]{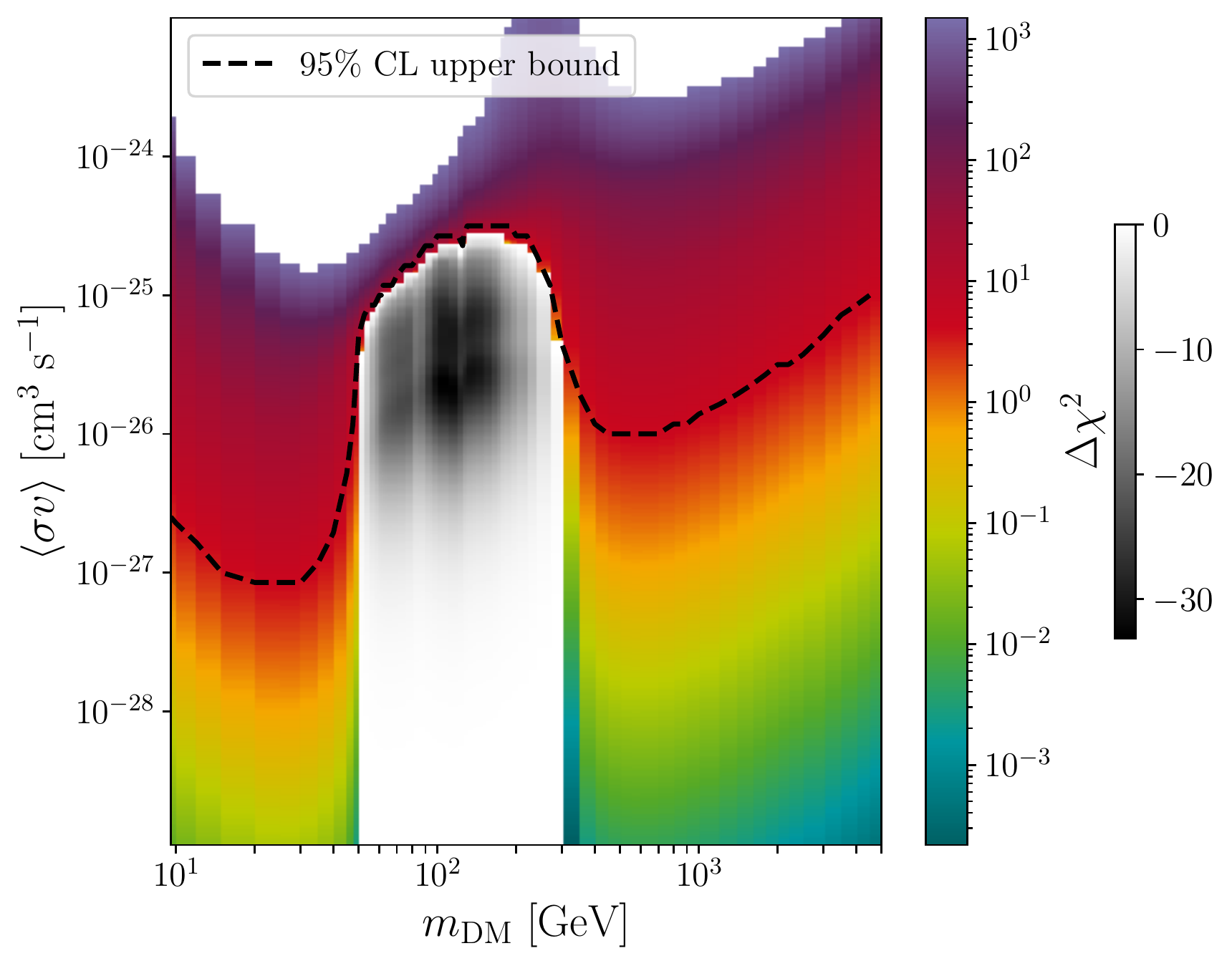}
	\end{minipage}
	\caption{\textit{Left:} Mass dependence of branching fractions of $S S \rightarrow \text{SM} \text{ SM}$ in the SSDM model for $\lambda_\mathrm{HS}$ fixed by the relic density requirement. \textit{Right:} Marginal $\chi^2$ distribution of the $\langle \sigma v \rangle - m_\text{DM}$ parameter space in the SSDM model.}
	\label{img:bounds_SSDM}
\end{figure}

The DM phenomenology of the SSDM is fully specified by the mass of the DM particle, $m_S=m_{\rm DM}$, and the strength of the coupling between the DM and Higgs particle, $\lambda_{H\!S}$. Below the Higgs-pair threshold, $m_S < m_h$, DM annihilation proceeds through $s$-channel Higgs exchange only, and the relative weight of the different SM final states is determined by the SM Higgs branching ratios, independent of the Higgs-scalar coupling $\lambda_{H\!S}$. Above the Higgs-pair threshold, $m_S \ge m_h$, the $hh$ final state opens up. The strength of the annihilation into Higgs pairs, as compared to $W$, $Z$ or top-quark pairs, depends on the size of the Higgs-scalar coupling. For our specific analysis we require that the SSDM provide the correct DM relic density, $\Omega h^2 = 0.1198\pm 0.0015$~\cite{Ade:2015xua}, which in turn determines the size of $\lambda_{H\!S}$ for any given DM mass $m_S$. The corresponding branching fractions for DM annihilation within the SSDM are shown in figure~\ref{img:bounds_SSDM} (left panel) as a function of the DM mass. 

Using the ANN we analyse the $\Delta\chi^2$ distribution of the model, marginalising over propagation uncertainties as described in section~\ref{sec:marg_importance}. The result is shown in figure~\ref{img:bounds_SSDM} (right panel). Comparing figure~\ref{img:bounds_SSDM} with the analogous result for the single annihilation channel into $b\bar{b}$, figure~\ref{img:bounds_bb} (right panel), we observe a similar overall shape of the $\Delta\chi^2$ distribution. 

For light DM the SSDM annihilates dominantly into bottom final states, so one expects results that are very similar to the case of the single $b\bar{b}$ channel. However, for the smallest DM masses that we consider ($m_\chi \approx 10 \, \mathrm{GeV}$) we find that the constraints become considerably stronger when including even a sub-dominant contribution from $c\bar{c}$. The reason is that in this mass range, most antiprotons resulting from annihilation into bottom quarks have energies below $5\,\mathrm{GeV}$ and do therefore not give a contribution in our fits. Annihilation into charm quarks, on the other hand, can give rise to more energetic antiprotons, leading to stronger constraints. For DM masses above about $50 \, \mathrm{GeV}$, a variety of SM final states contributes in the SSDM, including in particular $WW$, $hh$ and $ZZ$.  However, as shown in Ref.~\cite{Cuoco:2017iax}, the limits for heavy DM are similar for these final states and for annihilation into bottom quarks, so that the overall constraints for the SSDM are comparable to those for annihilation into bottom quarks only. 

\section{Conclusions}
\label{sec:conclusions}

The analysis of cosmic ray (CR) antiprotons is a powerful method for the indirect detection of dark matter (DM). The accurate experimental measurements, in particular from AMS-02, allow to probe DM annihilation cross sections close to the value predicted by thermal freeze-out for a wide range of DM masses. However, a precise description of CR propagation through the Galaxy is required to exploit the potential of the experimental data. The propagation models depend on a large number of parameters, and the  standard numerical simulation tools, such as \textsc{Galprop}, are computationally expensive. Therefore, global analyses of generic models of DM can only be carried out with an immense computational effort, if at all. 

In this work we have developed an artificial neural network (ANN) that allows extremely fast and accurate predictions of the cosmic ray flux for generic DM models. Specifically, we have employed recurrent neural networks (RNNs) to predict the CR energy spectrum. RNNs are particularly well suited to learn the correlations between the fluxes contained in neighbouring energy bins. Additional improvements in performance are achieved by grouping input parameters that have similar physical origin and by performing a suitable rescaling of the output spectra. 

We have trained the ANN with a large set of antiproton fluxes simulated with \textsc{Galprop}, where the propagation parameters have been chosen to be broadly compatible with the most recent AMS-02 data, and a generic parametrisation of the dark matter model in terms of the DM mass and the branching fractions for the annihilation into various 
Standard Model final states. We emphasise that the contribution of different DM models to the antiproton flux only has a marginal impact on the preferred range of the propagation parameters. It is therefore possible to focus the training of the ANN on the relevant range of propagation parameters without specifying the details of the DM model in advance. We have validated the performance and accuracy of the network by comparing both the predicted antiproton fluxes and the resulting AMS-02 likelihoods to the ones obtained from explicit \textsc{Galprop} simulations for a range of different propagation and DM model parameters.

We have then used the neural network predictions to test specific DM models against current AMS-02 data. We have focused on the DM parameter space and treated the propagation parameters as nuisance parameters by calculating both the corresponding profile and marginalised likelihoods. While the former approach requires an explicit restriction of the parameter space to the regions where the ANN has been sufficiently trained, this requirement can be automatically fulfilled in the latter case by employing importance sampling. Comparing the ANN to \textsc{Galprop} we find a speed-up in runtime of about two (three) orders of magnitude when using profiling (importance sampling).

For DM annihilation into bottom quarks we have obtained results that are consistent with previous studies based on simulations and a profile likelihood approach. We find more stringent bounds on the DM parameter space when using the marginalised likelihood; here a thermal cross section can be excluded for DM annihilating fully into bottom quarks for DM  masses in the range between approximately 300~GeV and 2~TeV. 
To illustrate the flexibility of our approach, we have also used the ANN to derive constraints on scalar singlet DM, for which DM annihilation results in a variety of Standard Model final states with branching fractions that depend strongly on the DM mass. 

The ANN developed in this work, and the corresponding method for efficient training, can {also be used to study more closely the potential DM interpretation of the antiproton excess around 20 GeV, for example regarding the impact of correlations in AMS-02 data. Moreover, it can} be easily extended to alternative propagation models and can be applied to a wide class of DM scenarios.
It will thus be possible to fully exploit the potential of current and future cosmic-ray data in global analyses of general DM models. {In future work a transformation of the ANNs into Bayesian neural networks can be incorporated in the analysis. With this step, additional more in-depth studies of the uncertainties of the network predictions will be possible.} The fully trained networks together with a suitable user interface are publicly available as \textsc{DarkRayNet} at \url{https://github.com/kathrinnp/DarkRayNet}.

\acknowledgments

We thank Thorben Finke and Christoph Weniger for discussions, Alessandro Cuoco and Jan Heisig for helpful comments on the manuscript and Sven Guenther for testing \textsc{DarkRayNet}. F.K.\ is  supported by  the  Deutsche Forschungsgemeinschaft (DFG) through the Emmy Noether Grant No.\ KA 4662/1-1.
M.Ko.\ is partially supported by the Swedish National Space Agency under contract 117/19 and the European Research Council under grant 742104.
Simulations and ANN training were performed with computing resources granted by RWTH Aachen University under project jara0184 and rwth0754.

\begin{appendix}

\section{Predicting proton and helium spectra}
\label{app:p_and_He}

\begin{figure}[t]
	\centering
	\includegraphics[width = 0.8\textwidth]{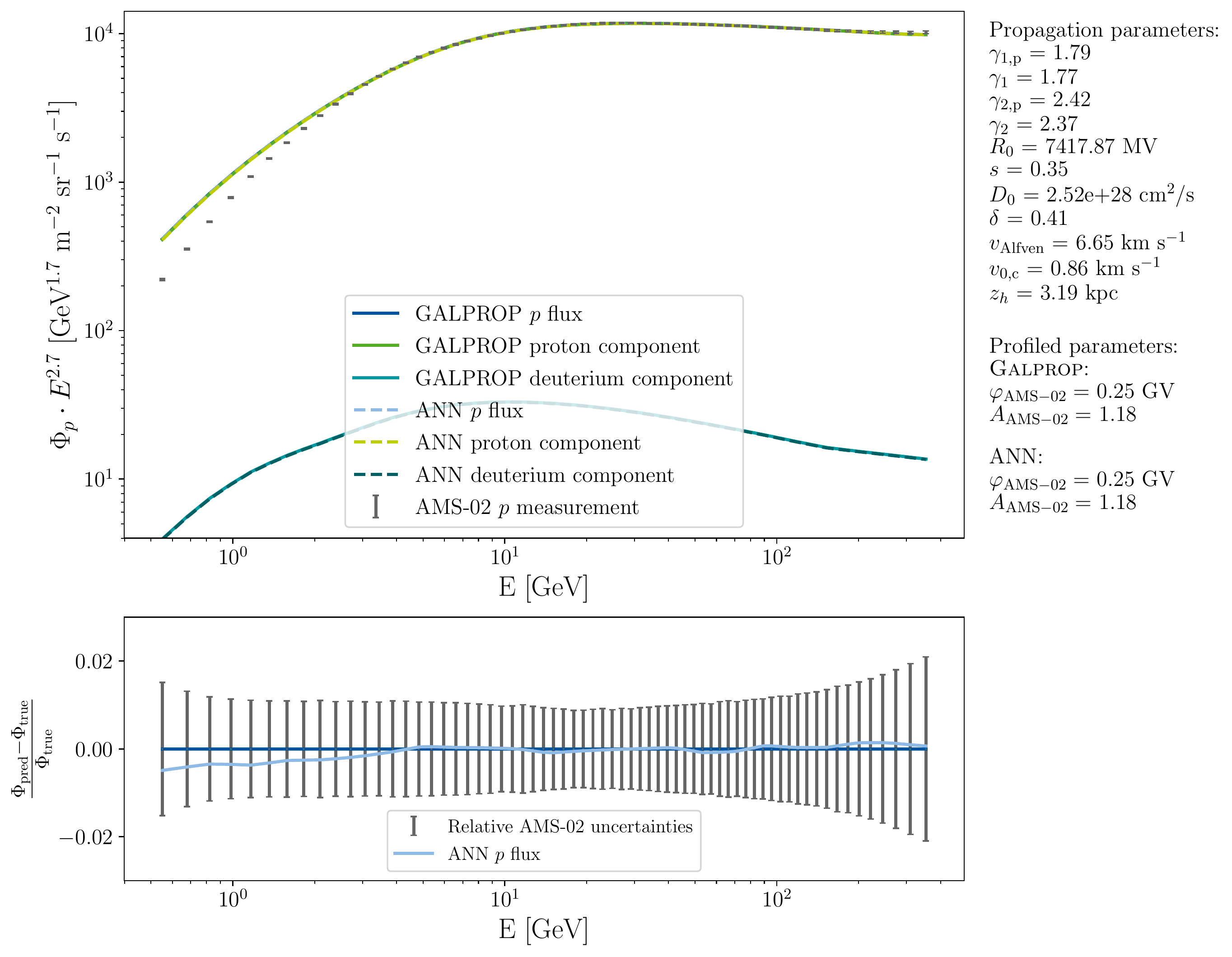}
	\caption{Exemplary comparison of the simulated versus predicted protons flux of the individual components protons and Deuterium and the combination of both where the listed parameters and simulated fluxes are randomly sampled from the test set. Each component of the neural network flux is predicted by the individual networks. Lower panel as figure~\ref{img:example_fluxes}.}
	\label{img:example_fluxes_p}
\end{figure}

\begin{figure}[t]
	\centering
	\includegraphics[width = 0.8\textwidth]{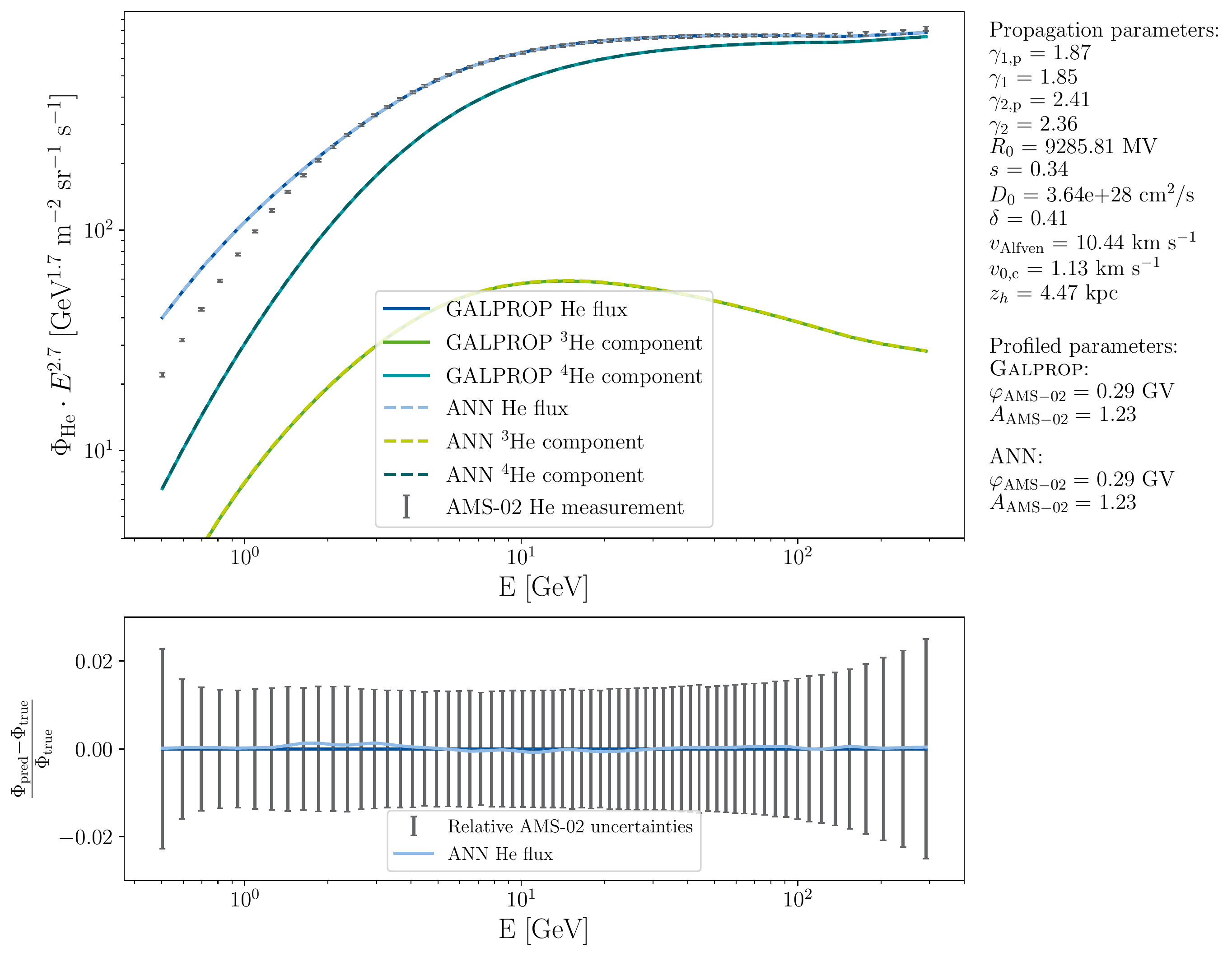}
	\caption{Exemplary comparison of the simulated versus predicted He flux of the individual components $^3$He and $^4$He and the combination of both where the listed parameters and simulated fluxes are randomly sampled from the test set. Each component of the neural network flux is predicted by the individual networks. Lower panel as figure~\ref{img:example_fluxes}.}
	\label{img:example_fluxes_He}
\end{figure}

When simulating the antiproton fluxes as described in section~\ref{sec:training_set} we can  also obtain the CR spectra of protons, deuterium, and helium ($^3$He and $^4$He) without significant additional computation costs due to the setup of \textsc{Galprop}. The task of modelling these spectra using an ANN is very comparable with the task fulfilled by the sNet. We have thus examined the ability of the sNet architecture (as described in sec.~\ref{sec:architectures}) to also accurately predict proton and helium spectra. The inputs of the sNet remain the same, but we have extended the length of the final output layer, to accommodate a wider energy range, appropriate for the proton and Helium AMS-02 and Voyager data. Using also the same training process (see sec.~\ref{sec:train_process}) we achieve a similar accuracy as for the secondary antiprotons, as each of the predictions deviates from the simulations only marginally with respect to the experimental uncertainties. In figures~\ref{img:example_fluxes_p} and \ref{img:example_fluxes_He} we show exemplary results for protons, resp. helium, and their individual components analogous to figure~\ref{img:example_fluxes}. 

\end{appendix}

\providecommand{\href}[2]{#2}\begingroup\raggedright\endgroup

\end{document}